\documentclass[10pt]{article}
\usepackage{euscript,amssymb,epsfig}
\usepackage{amsfonts,latexsym}
\usepackage{euscript,amssymb}
\usepackage{epsfig}

\def\RR{{\mathbb R}}

\catcode`\@=11

\def\vereq#1#2{\lower3pt\vbox{\baselineskip1.5pt \lineskip1.5pt
\ialign{$\m@th#1\hfill##\hfil$\crcr#2\crcr\sim\crcr}}}
\def\gtrsim{\mathrel{\mathpalette\vereq>}}

\def\Let@{\relax\iffalse{\fi\let\\=\cr\iffalse}\fi}
\def\vspace@{\def\vspace##1{\crcr\noalign{\vskip##1\relax}}}
\def\multilimits@{\bgroup\vspace@\Let@
 \baselineskip\fontdimen10 \scriptfont\tw@
 \advance\baselineskip\fontdimen12 \scriptfont\tw@
 \lineskip\thr@@\fontdimen8 \scriptfont\thr@@
 \lineskiplimit\lineskip
 \vbox\bgroup\ialign\bgroup\hfil$\m@th\scriptstyle{##}$\hfil\crcr}
\def\Sb{_\multilimits@}
\def\endSb{\crcr\egroup\egroup\egroup}
\def\Sp{^\multilimits@}

\newcommand{\be}[1]{\begin{equation}\label{#1}}
\newcommand{\ee}{\end{equation}}
\newcommand{\ba}[1]{\begin{eqnarray}\label{#1}}
\newcommand{\ea}{\end{eqnarray}}
\newcommand{\rf}[1]{(\ref{#1})}
\newcommand{\nn}{\nonumber}
\newcommand{\bmatrix}[1]{\left( \begin{array}{#1}}
\newcommand{\ematrix}{\end{array}\right)}

\newcommand{\sign}{ \mbox{\rm sign}\,}

\newlength{\indentedwidth}
\newdimen\mathindent
\indentedwidth=\mathindent

\newenvironment{indented}{\begin{indented}}{\end{indented}}
\newenvironment{varindent}[1]{\begin{varindent}{#1}}{\end{varindent}}
\def\indented{\list{}{\itemsep=0\p@\labelsep=0\p@\itemindent=0\p@
   \labelwidth=0\p@\leftmargin=\mathindent\topsep=0\p@\partopsep=0\p@
   \parsep=0\p@\listparindent=15\p@}\footnotesize\rm}

\def\varindent#1{\setlength{\varind}{#1}%
   \list{}{\itemsep=0\p@\labelsep=0\p@\itemindent=0\p@
   \labelwidth=0\p@\leftmargin=\varind\topsep=0\p@\partopsep=0\p@
   \parsep=0\p@\listparindent=15\p@}\footnotesize\rm}

\newcommand{\opensquare}{\mbox{$\rlap{$\sqcap$}\sqcup$}}
\def\etal{{\it et al\/}}

\begin{document}
\author{Uwe G\"unther$^a$\footnote{e-mail:
u.guenther@fz-rossendorf.de, present address: Research Center
Rossendorf, P.O. Box 510119, D-01314 Dresden, Germany}\, , Alexander
Zhuk$^{bc}$\footnote{e-mail: zhuk@paco.net}\, , Valdir B.
Bezerra$^c$\footnote{e-mail: valdir@fisica.ufpb.br}\, , and Carlos
Romero$^c$\footnote{e-mail: cromero@fisica.ufpb.br}
\\[2ex]
$^a$ Gravitationsprojekt, Mathematische Physik I,\\ Institut f\"ur
Mathematik, Universit\"at Potsdam,\\ Am Neuen Palais 10, PF
601553, D-14415 Potsdam, Germany\\[1ex]
$^b$ Department of Physics, University of Odessa,\\ 2 Dvoryanskaya
St., Odessa 65100, Ukraine\\[1ex]
$^c$ Departamento de F\'{\i}sica, Universidade Federal de
Para\'{\i}ba\\
C.Postal 5008, Jo$\tilde{\mbox{a}}$o Pessoa, PB, 58059-970, Brazil}
\title{AdS and stabilized extra dimensions in  multidimensional
gravitational models with nonlinear scalar curvature terms $R^{-1}$
and $R^4$}

\date{09 June 2005}
%
%
\maketitle

\noindent

\begin{abstract}
We study multidimensional gravitational models with scalar curvature
nonlinearities of the type $R^{-1}$ and $R^4$. It is assumed that
the corresponding higher dimensional spacetime manifolds undergo a
spontaneous compactification to  manifolds with warped product
structure. Special attention is paid to the stability of the
extra-dimensional factor spaces. It is shown that for certain
parameter regions the systems allow for a freezing stabilization of
these spaces. In particular, we find for the $R^{-1}-$model that
configurations with stabilized extra dimensions do not provide a
late-time acceleration (they are AdS), whereas the solution branch
which allows for accelerated expansion (the dS branch) is
incompatible with stabilized factor spaces. In the case of the
$R^4-$model, we obtain that the stability region in parameter space
depends on the total dimension $D=\dim(M)$ of the higher dimensional
spacetime M. For $D>8$ the stability region consists of a single
(absolutely stable) sector which is shielded from a conformal
singularity (and an antigravity sector beyond it) by a potential
barrier of infinite height and width. This sector is smoothly
connected with the stability region of a curvature-linear model. For
$D<8$ an additional (metastable) sector exists which is separated
from the conformal singularity by a potential barrier of finite
height and width so that systems in this sector are prone to
collapse into the conformal singularity. This second sector is not
smoothly connected with the first (absolutely stable) one. Several
limiting cases and the possibility for inflation are discussed for
the $R^4-$model.
\end{abstract}

\bigskip

\hspace*{0.950cm} PACS numbers: 04.50.+h, 11.25.Mj, 98.80.Jk

\section{\label{intro}Introduction}

Distance measurements of type Ia supernovas (SNe,Ia)
\cite{supernova} as well as cosmic microwave background (CMB)
anisotropy measurements \cite{cmb1} performed during the last
years give strong evidence for the existence of dark energy --- a
smooth energy density with negative pressure which causes an
accelerated expansion of the Universe at present time. This late
time acceleration stage should have started approximately $5$
billion years ago and represents a second acceleration epoch after
inflation which lasted for $10^{-35}$ seconds immediately after
Big Bang and ended $13.7$ billion years ago.

The challenge to theoretical cosmology consists in finding a
natural explanation of inflation and late time acceleration (dark
energy) within the framework of string theory/M-theory or loop
quantum gravity. Scenarios which address one or both of these
issues are, e.g., string (pre-bigbang) cosmology
\cite{string_cosmology}, a large number of brane world scenarios
(brane inflation \cite{on-brane-inflation}, the ekpyrotic
\cite{ekpyr} and born-again Universe \cite{born-again} scenario,
late-time acceleration via S-branes \cite{SUGRA1}), as well as the
string theory/M-theory scenarios with flux compactifications
\cite{multi-1,KKLT,flux-1,multi-2,flux-2} and the recent setups of
loop quantum  cosmology \cite{loop-2}.

A different starting point for explaining the late time
acceleration was taken in Refs. \cite{1/R-1,NO1}, where it was
shown that purely gravitational modifications of the (curvature
linear) Einstein-Hilbert action by including curvature nonlinear
terms of the type $\bar R^{-1}$, $\bar R^{-n}$ could induce a
positive effective cosmological constant, and with it an
accelerated expansion. The corresponding phenomenological studies
were performed in four spacetime dimensions by transforming the
curvature-nonlinear theory into an equivalent curvature-linear
theory with the nonlinearity degrees of freedom carried by an
additional dynamical scalar field. (The used technique had been
developed in earlier work since the 1980s
\cite{Kerner,BC,Maeda,EKOY}.) A natural question which arises with
regard to this phenomenological approach is of whether it can
naturally follow as low-energy limit from some M-theory setup or
--- looking from down-side up --- of whether it has a physically
viable phenomenological extension to higher dimensions.

In the present paper we take the latter (phenomenological) point of
view and study higher dimensional extensions of purely gravitational
non-Einsteinian models with scalar curvature nonlinearities of the
types $\bar R^{-1}$ and $\bar R^4$. Special emphasis will be laid on
finding parameter regions (regions in moduli space) which ensure the
existence of at least one minimum of the effective potential for the
volume moduli of the internal spaces and which by this way allow for
their stabilization. The latter fact is of special importance
because the extra-dimensional space components should be static or
nearly static at least from the time of primordial nucleosynthesis
(otherwise the fundamental physical constants would vary, see e.g.
\cite{GZ(CQG2001),Cline-1}\footnotetext[1]{First discussions of this
subject date back to Refs. \cite{horwath}.}, and observational
bounds on the variation of the fine-structure constant could be
violated \cite{GSZ}). During the last few years, problems of volume
moduli stabilization have been studied, e.g., for models with large
extra dimensions (Arkani-Hamed--Dimopoulos--Dvali (ADD) setups
\cite{ADD-Ant}) \cite{ADD-stab}\footnotetext[2]{See also
\cite{Ant-2}.}, as well as (more recently) for M-theory scenarios
with flux compactifications
\cite{KKLT,flux-1,multi-2,flux-2,stab-6,stab-7} and for brane-gas
systems \cite{kaya}.

Here we will mainly follow our earlier work on this subject
\cite{GSZ,GZ(PRD1997),GZ(CQG1),GZ(PRD2000),GMZ1,GMZ2}, where the
stabilization of extra dimensions was studied for
$(D_0+D')-$dimensional bulk spacetimes with a product topology.
The corresponding product manifolds are constructed from Einstein
spaces $M_i$ with scale (warp) factors which depend only on the
coordinates of the external $D_0-$dimensional spacetime $M_0$ (the
ansatz resembles a zero mode approximation in Kaluza-Klein
formalism). As a consequence, the excitations of the scale factors
(conformal excitations/excitations of the volume moduli) have the
form of massive scalar fields (gravitational excitons/radions)
living in the external spacetime. Stabilized volume moduli will
correspond to positive eigenvalues of the mass matrix of these
scalar fields, unstable configurations to tachyonic excitations.

The present work can be understood as a direct continuation of our
earlier investigation on volume moduli stabilization in $\bar
R^2-$models of purely geometrical type \cite{GMZ1} as well as with
magnetic (solitonic, Freund-Rubin-type) form fields living in the
extra dimensions (flux field compactifications) \cite{GMZ2}. Its
key results can be summarized as follows:
\begin{itemize}
\item A straight forward extension of a four-dimensional purely
geometrical $\bar R^{-1}-$model to higher dimensions with
subsequent dimensional reduction cannot simultaneously provide a
late time acceleration and a stabilization of the extra
dimensions. A late time acceleration is only possible for a
solution branch which has a positive definite maximum of the
effective potential, i.e. a positive definite effective
cosmological constant, and not a negative definite minimum as it
would be required for a stabilization of the internal factor space
components. This means that other, more sophisticated, extension
scenarios would be needed to reach both goals simultaneously.
\item In contrast to the $\bar R^2-$models of Refs.
\cite{GMZ1,GMZ2}, the $\bar R^4-$model shows a rich substructure
of the stability region in parameter (moduli) space which
crucially depends on the total dimension $D=D_0+D'$ of the bulk
spacetime. There exists one stability sector which is present for
all dimensions $D\ge D_0+2$ and which smoothly tends to the
stability sector of an $\bar R-$linear model when the $\bar
R-$nonlinearity is switched off. This sector is shielded from the
(probably unphysical) antigravity sector of the theory by a
potential barrier of infinite height and width, and hence, it will
be absolutely stable with regard to transitions of the system into
the antigravity sector. Apart from this sector of absolute
stability, there exists a second sector for total dimensions $D<8$
which is separated from a conformal singularity and an antigravity
sector beyond it by a potential barrier of finite height and
width. Systems in this sector will only be metastable and prone to
collapse into the conformal singularity and the antigravity
sector. The metastable sector is separated from the stability
sector of the $\bar R-$linear model by an essential singularity in
$\bar R$ and the effective potential $U_{eff}$ (when $\bar R$ and
$U_{eff}$ are considered as functions over a parameter subspace).
\item In the analyzed purely geometrical $\bar R-$nonlinear models,
a stabilization of the internal factor spaces is necessarily
connected with an $AdS$ structure of the external spacetime. In
order to uplift the $AdS$ to an $dS$ with positive effective
cosmological constant and the possibility for late time
acceleration, additional matter fields should be included into the
model (e.g., flux fields as in the $\bar R^2-$model of Ref.
\cite{GMZ2}).
\end{itemize}

The paper is structured as follows. In section \ref{setup}, we
present a brief technical outline of the transformation from a
non-Einsteinian purely gravitational model with general scalar
curvature nonlinearity of the type $f(\bar R)$ to an equivalent
curvature-linear model with additional nonlinearity carrying
scalar field. Afterwards, we derive in Sec. \ref{freezing}
criteria which ensure the existence of at least one minimum for
the effective potential of the internal space scale factors
(volume moduli). These criteria are then used in sections
\ref{first model} and \ref{third model} to obtain the regions in
parameter (moduli) space which allow for a freezing stabilization
of the scale factors in models with scalar curvature
nonlinearities of type $\bar R^{-1}$ and $\bar R^4$. The main
results are summarized in the concluding Sec. \ref{conclu}.

\section{\label{setup}General setup}

We consider a $D= (4+D^{\prime})-$dimensional nonlinear pure
gravitational theory with action functional
\be{1.1} S = \frac {1}{2\kappa^2_D}\int_M d^Dx \sqrt{|\bar g|}
f(\bar R)\; , \ee
where $f(\bar R)$ is an arbitrary smooth function with mass
dimension $\mathcal{O}(m^2)$ \ ($m$ has the unit of mass) of a
scalar curvature $\bar R = R[\bar g]$ constructed from the
$D-$dimensional metric $\bar g_{ab}\; (a,b = 1,\ldots,D)$.
$D^{\prime}$ is the number of extra dimensions and $\kappa^2_D $
denotes the $D-$dimensional gravitational constant which is
 connected with the fundamental mass scale
$M_{*(4+D^{\prime})}$ and the surface area $S_{D-1}=2\pi
^{(D-1)/2}/\Gamma [(D-1)/2]$ of a unit sphere in $D-1$ dimensions
by the relation \cite{add1,GZ-mg10,GZ-n1}
\be{1.2}  \kappa^2_D = 2S_{D-1} /
M_{*(4+D^{\prime})}^{2+D^{\prime}}.\ee
The equation of motion for the theory \rf{1.1} reads (see, e.g.,
Refs. \cite{Kerner,BC,Maeda})
\be{1.3} f^{\prime }\bar R_{ab} -\frac12 f\, \bar g_{ab} - \bar
\nabla_a \bar \nabla_b f^{\prime } + \bar g_{ab} \bar{\opensquare
} f^{\prime } = 0\;  \ee
and has as trace
\be{1.5} (D-1)\bar{\opensquare } f^{\prime } = \frac{D}{2} f
-f^{\prime }\bar R\; . \ee
We use the notations $\; \bar \nabla_a$ and $\bar{\opensquare}$
for the covariant derivative and the Laplacian with respect to the
metric $\bar g_{ab}$, as well as the abbreviations $ f^{\prime }
=df/d\bar R$, $\; \bar R_{ab} = R_{ab}[\bar g]$.

Before we endow the metric of the  pure gravity theory \rf{1.1}
with explicit structure, we recall that this $\bar R-$nonlinear
theory is equivalent to a theory which is linear in another scalar
curvature $R$ but which contains an additional self-interacting
scalar field. According to standard techniques
\cite{Kerner,BC,Maeda}, the corresponding $R-$linear theory has
the action functional:
\be{1.6} S = \frac{1}{2\kappa^2_D} \int_M d^D x \sqrt{|g|} \left[
R[g] - g^{ab} \phi_{,a} \phi_{,b} - 2 U(\phi )\right]\; , \ee
where
\be{1.7} f'(\bar R) = \frac {df}{d \bar R} := e^{A \phi} > 0\; ,\quad A :=
\sqrt{\frac{D-2}{D-1}}\;  ,\ee
and where the self-interaction potential $U(\phi )$ of the scalar
field $\phi$ is given by
\ba{1.8-1} U(\phi ) &=&
 \frac12 \left(f'\right)^{-D/(D-2)} \left[\; \bar R f' - f\right]\;
 ,\label{1.8a}\\
&=& \frac12 e^{- B \phi} \left[\; \bar R (\phi )e^{A \phi } -
f\left( \bar R (\phi )\right) \right]\; , \quad B := \frac
{D}{\sqrt{(D-2)(D-1)}}\, .\label{1.8}
\ea
The metrics $g_{ab}$, $\bar g_{ab}$ and the scalar curvatures $R$,
$\bar R$ of the two theories \rf{1.1} and \rf{1.6} are conformally
connected by the relations
\be{1.9} g_{ab} = \Omega^2 \bar g_{ab} = \left[ f'(\bar
R)\right]^{2/(D-2)}\bar g_{ab}\;  \ee
and
\be{1.10} R = (f')^{2/(2-D)}\left\{ \bar R +\frac{D-1}{D-2}
(f')^{-2} \bar g^{ab}\partial_a f'\partial_b f'-
2\frac{D-1}{D-2}(f')^{-1} \bar {\opensquare} f'\right\}\;  \ee
via the scalar field $\phi=\ln[f'(\bar R)]/A$. This scalar field
$\phi$ carries the nonlinearity degrees of freedom in $\bar R$ of
the original theory, and for brevity we call it the nonlinearity
field.

As next, we assume that the D-dimensional bulk space-time $M$
undergoes a spontaneous compactification\footnotetext[3]{For a
discussion of possible decompactification scenarios we refer to the
recent work \cite{decomp-1}.} to a warped product
 manifold
\be{1.18} M = M_0 \times M_1 \times \ldots \times M_n \ee
with  metric
\be{1.19} \bar g=\bar g_{ab}(X)dX^a\otimes dX^b=\bar
g^{(0)}+\sum_{i=1}^ne^{2\bar {\beta} ^i(x)}g^{(i)}\; . \ee
The coordinates on the $(D_0=d_0+1)-$dimensional manifold $M_0 $
(usually interpreted as our observable $(D_0=4)-$dimensional
Universe) are denoted by $x$ and the corresponding metric by
\be{1.20} \bar g^{(0)}=\bar g_{\mu \nu }^{(0)}(x)dx^\mu \otimes
dx^\nu\; . \ee
For simplicity, we choose the internal factor manifolds $M_i$ as
$d_i-$dimensional Einstein spaces with metrics
$g^{(i)}=g^{(i)}_{m_in_i}(y_i)dy_i^{m_i}\otimes dy_i^{n_i},$ so
that the relations
\be{1.21} R_{m_in_i}\left[ g^{(i)}\right] =\lambda
^ig_{m_in_i}^{(i)},\qquad m_i,n_i=1,\ldots ,d_i \ee
and
\be{1.22} R\left[ g^{(i)}\right] =\lambda ^id_i\equiv R_i
\;  \ee hold.
The specific metric ansatz \rf{1.19} leads to a scalar curvature
$\bar R$ which depends only on the coordinates $x$ of the external
space: $\bar R[\bar g] = \bar R(x)$. Correspondingly, also the
nonlinearity field $\phi$ depends on $x$ only: $\phi = \phi (x)$.

Passing from the $\bar R-$nonlinear theory \rf{1.1} to the
equivalent $R-$linear theory \rf{1.6} the metric \rf{1.19}
undergoes the conformal transformation $\bar g \mapsto g$ [see
relation \rf{1.9}]
\be{1.23} g = \Omega^2 \bar g = \left( e^{A \phi
}\right)^{2/(D-2)} \bar g\: := g^{(0)}+\sum_{i=1}^ne^{2
\beta^i(x)}g^{(i)}\;  \ee
with
\be{1.24} g^{(0)}_{\mu \nu} := \left( e^{A \phi}\right)^{2/(D-2)}
\bar g^{(0)}_{\mu \nu}\; , \quad
 \beta^i := \bar {\beta} ^i + \frac{A}{D-2} \phi\; . \ee

\section{Freezing stabilization\label{freezing}}

The main subject of our subsequent considerations will be the
stabilization of the internal space components. A strong argument
in favor of stabilized or almost stabilized internal space scale
factors $\bar {\beta} ^i(x)$, at the present evolution stage of
the Universe, is given by the intimate relation between variations
of these scale factors and those of the fine-structure constant
$\alpha$ \cite{GSZ}. The strong restrictions on
$\alpha-$variations in the currently observable part of the
Universe \cite{alpha-var} imply a correspondingly strong
restriction on these scale factor variations \cite{GSZ}. For this
reason, we will concentrate below on the derivation of criteria
which will ensure a freezing stabilization of the scale factors.
Extending earlier discussions of models with $\bar R^2$ scalar
curvature nonlinearities \cite{GMZ1,GMZ2} we will investigate here
models of the nonlinearity types $\bar{R}^{-1}$ and $\bar{R}^4$.

In Ref. \cite{GZ(PRD2000)} it was shown that for models with a
warped product structure \rf{1.19} of the bulk spacetime $M$ and a
minimally coupled scalar field living on this spacetime, the
stabilization of the internal space components requires a
simultaneous freezing of the scalar field. Here we expect a
similar situation with simultaneous freezing stabilization of the
scale factors ${\beta} ^i(x)$ and the nonlinearity field
$\phi(x)$. According to \rf{1.24}, this will also imply a
stabilization of the scale factors $\bar {\beta} ^i(x)$ of the
original nonlinear model.

In general, the model will allow for several stable scale factor
configurations (minima in the landscape over the space of volume
moduli). We choose one of them\footnotetext[4]{Although the toy
model ansatz \rf{1.1} is highly oversimplified and far from a
realistic model, we can roughly think of the chosen minimum, e.g.,
as that one which we expect to correspond to the current evolution
stage of our observable Universe.}, denote the corresponding scale
factors as $\beta^i_0$, and work further on with the deviations
\be{1.25} \hat \beta^i (x)= \beta^{i}(x) - \beta^{i}_0\ee
as the dynamical fields\footnotetext[5]{For simplicity, we work here
with
 stabilized scale factors $\beta^i_0$ which we assume as homogeneous and constant. In
general, one can split the scale factors $\beta^i(x)$, e.g., into
a coherent scale factor background $\beta^i_0(x)$ and non-coherent
scale factor fluctuations $\hat \beta^i (x)= \beta^{i}(x) -
\beta^{i}_0(x)$ over this background \cite{GZ(CQG1)}. }. After dimensional
reduction of the action functional \rf{1.6} we pass from the
intermediate Brans-Dicke frame to the Einstein frame via a
conformal transformation
\begin{equation}
\label{1.26}g_{\mu \nu }^{(0)}=\hat{\Omega} ^2\hat g_{\mu \nu
}^{(0)} =\left( \prod^{n}_{i=1} e^{d_i\hat \beta ^i}
\right)^{-2/(D_0-2)} \hat g_{\mu \nu }^{(0)}\,
\end{equation}
with respect to the scale factor deviations $\hat \beta^i (x)$
\cite{GMZ1,GMZ2,GZ-mg10}. As result we arrive at the following
action
\be{1.27} S = \frac 1{2\kappa
_{D_0}^2}\int\limits_{M_0}d^{D_0}x\sqrt{|\hat g^{(0)}|}\left\{ \hat
R\left[ \hat g^{(0)}\right] -\bar G_{ij}\hat g^{(0)\mu \nu }\partial
_\mu\hat \beta ^i\,\partial _\nu \hat \beta ^j - \hat g^{(0)\mu \nu
}\partial _\mu\phi \,\partial _\nu\phi -2U_{eff}\right\} \, ,
\ee
which contains the scale factor offsets $\beta^i_0$ through the
total internal space volume
\be{1.29} V_{D'} \equiv V_I\times v_0\equiv
\prod^{n}_{i=1}\int_{M_i}d^{d_i}y \sqrt{|g^{(i)}|}\times
\prod^{n}_{i=1}e^{d_i\beta^{i}_0}\ee
in the definition of the effective gravitational constant $\kappa
_{D_0}^2$ of the dimensionally reduced theory
\be{1.28}\kappa _{(D_0=4)}^2=\kappa_D ^2/V_{D'}=
8\pi/M^2_{4}\quad \Longrightarrow \quad M_{4}^2 =
\frac{4\pi}{S_{D-1}}V_{D'} M_{*(4+D^{\prime})}^{2+D^{\prime}}.\ee
Obviously, at the present evolution stage of the Universe, the
internal space components should have a total volume which would
yield a four-dimensional mass scale of order of the Planck mass
$M_{(4)} = M_{Pl}$. The tensor components of the midisuperspace
metric (target space metric on $\RR _T^n$) $\bar G_{ij}\
(i,j=1,\ldots ,n)$, its inverse metric $\bar G^{ij} $ and the
effective potential are given as \cite{IMZ,RZ}
\begin{equation}
\label{1.30}\bar G_{ij}=d_i\delta _{ij}+\frac 1{D_0-2}d_id_j\, ,
\quad \bar G^{ij}=\frac{\delta ^{ij}}{d_i}+\frac 1{2-D}\, .
\end{equation}
The effective potential has the explicit form
\begin{equation}
\label{1.31}U_{eff} ( \hat \beta , \phi ) ={\left(
\prod_{i=1}^ne^{d_i\hat \beta ^i}\right) }^{-\frac 2{D_0-2}}\left[
-\frac 12\sum_{i=1}^n\hat R_{i}e^{-2\hat \beta ^i}+ U(\phi )
\right] \, ,
\end{equation}
where we abbreviated
\be{1.31a} \hat R_{i} := R_i
\exp{(-2\beta^{i}_{0})}.
\ee
For completeness, we note that the original metric $\bar g_{ab}$
of the $\bar R-$nonlinear model and the final Einstein frame
metric $\hat g_{\mu \nu}$ of the dimensionally reduced model are
connected by the relation
\be{1.31b}
\bar g_{ab}=
\left(f'\right)^{-\frac{2}{D-2}}\left[\left(\prod_{i=1}^n
e^{d_i\hat \beta^i}\right)^{-\frac{2}{D_0-2}} \hat{g}_{\mu
\nu}^{(0)} +\sum_{i=1}^n e^{2\beta^i}g^{(i)}\right]
\ee
which up to the nonlinearity induced conformal factor
$\left(f'\right)^{-2/(D-2)}$ takes (for scale factors depending
only on the time coordinate) a similar form like in the recently
analyzed  cosmological $S-$brane models of Refs.
\cite{SUGRA1,ivash-2}.

A freezing stabilization of the internal spaces will be
achieved\footnotetext[6]{An alternative stabilization mechanism can
consist, e.g., in the recently proposed dynamical stabilization in
the vicinity of enhanced symmetry points \cite{ESP}. In our present
discussion we will not analyze such scenarios.} if the effective
potential has at least one minimum with respect to the fields $\hat
\beta^i(x)$. Assuming, without loss of generality, that one of the
minima  is located at $\beta^{i} = \beta^{i}_{0} \Rightarrow \hat
\beta^{i} = 0$, we get the extremum condition:
\be{1.32} \left.\frac{\partial U_{eff}}{\partial \hat
\beta^{i}}\right|_{\hat \beta =0} =0 \Longrightarrow \hat R_{i} =
\frac{d_i}{D_0-2}\left( -\sum_{j=1}^n \hat R_{j} +2 U(\phi)
\right)\, . \ee
{}From its structure (a constant on the l.h.s. and a dynamical
function of $\phi (x)$ on the r.h.s) it follows that a
stabilization of the internal space scale factors can only occur
when the nonlinearity field $\phi (x)$ is stabilized as well. In
our freezing scenario this will require a minimum with respect to
$\phi$:
\be{1.33} \left. \frac{\partial U(\phi )}{\partial \phi
}\right|_{\phi_{0}} = 0\; \Longleftrightarrow\;  \left.
\frac{\partial U_{eff}}{\partial \phi}\right|_{\phi_{0}}  = 0\, .
\ee
We arrived at a stabilization problem, some of whose general
aspects have been analyzed already in Refs.
\cite{GZ(PRD1997),GZ(PRD2000),GMZ1,GMZ2}. For brevity we only
summarize the corresponding essentials as they will be needed for
more detailed discussions in the next sections.
\begin{enumerate}
\item \label{c1} Eq. \rf{1.32} implies that the scalar curvatures
$\hat R_i$ and with them the compactification scales
$e^{\beta^i_{0}}$ [see relation \rf{1.31a}] of the internal space
components are finely tuned
\be{1.34} \frac{\hat R_i}{d_i} = \frac{\hat
R_j}{d_j}\, , \quad i,j = 1,\ldots ,n \, . \ee
\item \label{c2} The masses of the normal mode excitations of the internal
space scale factors (gravitational excitons/radions) and of the
nonlinearity field $\phi$ near the minimum position are given as
\cite{GZ(PRD2000)}:
\ba{1.35} m_{1}^2 &=& \dots \; = m_{n}^2 = \, -\frac{4}{D-2}
U(\phi_{0})\,= -2\frac{\hat R_{i}}
{d_i} > 0\, , \label{1.35a} \\
&\phantom{-} & \nn\\
m_{\phi }^2 &:=& \left. \frac{d^2 U(\phi )}{d \phi^2}
\right|_{\phi_{0}}>0\, .\label{1.35b} \ea
\item \label{c3} The value of the effective potential at the minimum plays
the role of an effective 4D cosmological constant of the external
(our) spacetime $M_0$:
\be{1.36} \Lambda_{{eff}} :=\left. U_{eff}\vphantom{\int}
\right|_ {\hat \beta^i =0,\atop \phi = \phi_{0}}\; =\,
\frac{D_0-2}{D-2} U(\phi_{0})\, =\, \frac{D_0-2}{2}\frac{\hat
R_{i}}{d_i}\, .
\ee
\item \label{c4} Relation \rf{1.36} implies
\be{1.36c4}
\sign \Lambda_{eff} =\sign U(\phi_{0}) =\sign R_{i}\, .
\ee
Together with condition \rf{1.35} this shows that in a pure
geometrical model stable configurations can only exist for internal
spaces with negative curvature\footnotetext[7]{Negative constant
curvature spaces $M_i$ are compact if they have a quotient
structure: $M_i = H^{d_i}/\Gamma_i$, where $H^{d_i}$ and $\Gamma_i$
are hyperbolic spaces and their discrete isometry group,
respectively.} $R_{i} <0\; (i=1,\ldots ,n)$. Additionally, the
effective cosmological constant $\Lambda_{eff}$ as well as the
minimum of the potential $U(\phi )$ should be negative too:
\be{1.36c4a}
\Lambda_{eff} <0, \qquad U(\phi_{0})<0\, .
\ee
\item \label{c5} Eqs. \rf{1.29}, \rf{1.28}, \rf{1.31a} and \rf{1.34} - \rf{1.36} yield the following
scaling behavior of the minimum related model parameters under a
change of one of the offset scale factors $\beta_{0(1)}^i
\longrightarrow \beta_{0(2)}^i:=\ln (\lambda)\beta_{0(1)}^i$ as it
will be induced, e.g., by a change of the minimum value
$U(\phi_{0(1)})\longrightarrow U(\phi_{0(2)})$:
\ba{1.36c5a}
e^{\beta_{0(2)}^i}=\lambda e^{\beta_{0(1)}^i}&\quad
\Longrightarrow \quad & e^{\beta_{0(2)}^k}=\lambda
e^{\beta_{0(1)}^k} \quad \forall
k\, ,\label{1.36c5a1}\\
&& \frac{m^2_{k(1)}}{m^2_{k(2)}}=\frac{\hat R_{k(1)}}{\hat
R_{k(2)}}=\frac{\Lambda_{eff(1)}}{\Lambda_{eff(2)}}
=\frac{U(\phi_{0(1)})}{U(\phi_{0(2)})}=\lambda^2\, ,\label{1.36c5a2}\\
&& \frac{\kappa^2_{(D_0=4),(1)}}{\kappa^2_{(D_0=4),(2)}}
=\frac{V_{D' (2)}}{V_{D' (1)}}
=\lambda^{D'}=\left[\frac{U(\phi_{0(1)})}{U(\phi_{0(2)})}\right]^{D'/2}\label{1.36c5a3}.
\ea
\item \label{c6}
For a system which is almost stabilized at a freezing point $\phi
\approx \phi_0$, the function $f(\bar R)$ can be split into a
constant background and small deviations:
\be{1.36c6a}
f(\bar R) \approx c_1(\bar R - \bar R_0) + f(\bar R_0) \equiv c_1
\bar R + c_2\, ,
\ee
where $c_1 := f'(\bar R_0) = \exp(A \phi_0)$, \ $\bar R_0 \equiv
\bar R (\phi_0)$, and $-c_2/(2c_1)$ plays the role
of a cosmological constant. In the case of homogeneous and
isotropic spacetime manifolds, linear purely geometrical theories
with constant $\Lambda -$term necessarily imply an
anti-deSitter/deSitter geometry so that the manifolds are Einstein
spaces. Substitution of $ f(\bar R) \rightarrow c_1 \bar R + c_2$
into Eq. \rf{1.3} proves this fact directly
\be{1.12} \bar R_{ab} \longrightarrow
-\frac{1}{D-2}\frac{c_2}{c_1}\bar g_{ab}\quad \Longrightarrow
\quad \bar R \longrightarrow -\frac{D}{D-2}\frac{c_2}{c_1}\, .
\ee
\end{enumerate}
Plugging the potential $U(\phi)$ from Eq. \rf{1.8} into the
minimum conditions \rf{1.33}, \rf{1.35b}  yields with the help of
$\partial_\phi \bar R=A f'/f''$ the conditions
\ba{1.11}
\left.\frac{d U}{d\phi}\right|_{\phi_{0}} &=&
\left.\frac{A}{2(D-2)}\left(f'\right)^{-D/(D-2)}h\right|_{\phi_{0}}
= 0,\nn\\
h:&=& Df-2\bar R f', \quad \Longrightarrow \quad \
h(\phi_0) = 0\, ,\label{1.11a}\\
\left.\frac{d^2 U}{d\phi^2}\right|_{\phi_{0}} &=& \frac12 A
e^{(A-B)\phi_{0}} \left[ \partial_\phi \bar R + (A-B)\bar R
\right]_{\phi_{0}}\nn\\
&=& \left.\frac{1}{2(D-1)}
 \left(f'\right)^{-2/(D-2)} \frac{1}{f''}\partial_{\bar R} h \right|_{\phi_{0}}>0\, ,\label{1.11b}
\ea
where the last inequality can be reshaped into the suitable form
\be{1.11c}
\left.f''\partial_{\bar R}
h\right|_{\phi_{0}}=f''\left[(D-2)f'-2\bar R f''\right]_{\phi_0}>0
\; .
\ee
Furthermore, we find from Eq. \rf{1.11a}
\be{1.12-1}
U(\phi_0)=\frac{D-2}{2D}\left(f'\right)^{-\frac{2}{D-2}}\bar
R(\phi_0)
\ee
so that \rf{1.36c4a} leads to the additional restriction
\be{1.13-1}
\bar R(\phi_0)<0
\ee
at the extremum. Via the relation \rf{1.10} the stabilization
point curvatures of the $\bar R-$nonlinear and the $R-$linear
models are connected as
\be{1.13-1a}
R_0 \approx (f')^{2/(2-D)} \bar R_0.
\ee
Thus, as the extra dimensional scale factors approach their
stability position the bulk spacetime curvature asymptotically
(dynamically) tends to a negative constant value (see Eq.
\rf{1.13-1}). Because the effective cosmological constant is also
negative ($\Lambda_{eff} <0$), the homogeneous and isotropic
external $(D_0=4)-$dimensional spacetime is asymptotically
$\mbox{AdS}_{D_0}$. Together with the compact hyperbolic internal
spaces $M_i = H^{d_i}/\Gamma_i$ this results in a spontaneous
compactification scenario
\be{1.42} \mbox{AdS}_D\longrightarrow \mbox{AdS}_{D_0}\times
H^{d_1}/\Gamma_1 \times \dots \times H^{d_n}/\Gamma_n \; .\ee

In the next sections we will analyze the conditions \rf{1.34} -
\rf{1.11b}, \rf{1.13-1} on their compatibility with particular
scalar curvature nonlinearities $f(\bar R)$.

\section{The $R^{-1}-$model \label{first model}}

Recently it has been shown in Refs. \cite{1/R-1} that cosmological
models with a nonlinear scalar curvature term of the type $\bar
R^{-1}$ can provide a possible explanation of the observed late-time
acceleration of our Universe within a pure gravity setup. The
equivalent linearized model contains an effective potential with a
positive branch which can simulate a transient inflation-like
behavior in the sense of an effective dark energy. The corresponding
considerations have been performed mainly in four
dimensions\footnotetext[8]{A discussion of pro and contra of a
higher dimensional origin of $\bar R^{-1}$ terms can be found in
Ref. \cite{NO1}.}. Here we extend these analyses   to higher
dimensional models
--- assuming that the scalar curvature nonlinearity is of the same
form in all dimensions. We start from a nonlinear coupling of the
type:
\be{2.1} f(\bar R) = \bar R - \mu / \bar R\, , \quad \mu >0 \,
.\ee
In front of the $\bar R^{-1}-$term, the minus sign  is chosen,
because otherwise the potential $U(\phi )$ will have no extremum.

With the help of definition \rf{1.7}, we express the scalar
curvature $\bar R$ in terms of the nonlinearity field $\phi$ and
obtain two real-valued solution branches
\be{2.2} \bar R_\pm = \pm \sqrt{\mu} \left( e^{A\phi} -
1\right)^{-1/2}\, ,\quad \Longrightarrow \quad \phi >0
\ee
of the quadratic equation $f'(\bar R)=e^{A\phi}$.
The corresponding potentials
\be{2.3} U_\pm(\phi) = \pm \sqrt{\mu}\, e^{-B\phi }\,
\sqrt{e^{A\phi }-1}\, \ee
have extrema for curvatures [see Eq. \rf{1.11}]
\ba{2.4} \bar R_{0,\pm} &=&\pm \sqrt{\mu}\sqrt{\frac{D+2}{D-2}}\nn\\
e^{A\phi_0} &=& \frac{2B}{2B-A} = \frac{2D}{D+2}
> 1 \quad \mbox{for} \quad D \geq 3\, \ea
and take for these curvatures the values
\be{2.5}U_\pm(\phi_0) = \pm
\sqrt{\mu}\sqrt{\frac{D-2}{D+2}}\; e^{-B\phi_0} = \pm
\sqrt{\mu}\sqrt{\frac{D-2}{D+2}}\left( \frac{2D}{D+2}
\right)^{-D/(D-2)}\, .\ee
{}The stability defining second derivatives [Eq. \rf{1.11b}] at
the extrema \rf{2.4},
\ba{2.6} \left.  \partial^2_\phi U_\pm\right|_{\phi_{0}} &=&
\mp \sqrt{\mu} \frac{D}{D-1}\sqrt{\frac{D+2}{D-2}}\;
e^{B\phi_{0}}\nn\\ &=& \mp \sqrt{\mu}
\frac{D}{D-1}\sqrt{\frac{D+2}{D-2}} \left( \frac{2D}{D+2}
\right)^{-D/(D-2)}\, ,\ea
show that only the negative curvature branch $\bar R_-$ yields a
minimum with stable internal space components. The positive branch
has a maximum with $U_+(\phi_0)>0$. According to \rf{1.36c4} it
can provide an effective dark energy contribution with
$\Lambda_{eff}>0$, but due to its tachyonic behavior with
$\partial_\phi^2U(\phi_0)<0$ it cannot give stably frozen internal
dimensions. This means that the simplest extension of the
four-dimensional purely geometrical $\bar R^{-1}$ setup of Refs.
\cite{1/R-1} to higher dimensions is incompatible with a freezing
stabilization of the extra dimensions. A possible circumvention of
this behavior could consist in the  existence of different
nonlinearity types $f_i(\bar R_i)$ in different factor spaces
$M_i$ so that their dynamics can decouple one from the other. This
could allow for a freezing of the scale factors of the internal
spaces even in the case of a late-time acceleration with
$\Lambda_{eff}>0$. Another circumvention could consist in a
mechanism which prevents the dynamics of the internal spaces from
causing strong variations of the fine-structure constant $\alpha$.
The question of whether one of these schemes could work within a
physically realistic setup remains to be clarified.

Finally we note that in the minimum of the effective potential
$U_{eff}(\phi,\beta^i)$, which is provided by the negative
curvature branch $\bar R_(\phi)$, one finds excitation masses for
the gravexcitons/radions and the nonlinearity field (see Eqs.
\rf{1.35}, \rf{2.5} and \rf{2.6}) of order
\be{2.7} m_1 = \ldots = m_n \sim m_{\phi} \sim \mu^{1/4}\, .\ee
For the four-dimensional effective cosmological constant
$\Lambda_{eff}$ defined in \rf{1.36} one obtains in accordance
with Eq. \rf{2.5} \ $\Lambda_{eff} \sim - \sqrt{\mu}$.

\section{The $R^4-$model \label{third model}}

In this section we analyze a model with curvature-quartic
correction term of the type
\be{4.1} f(\bar R) = \bar R +\gamma \bar R^{4}
-2\Lambda_D\, .
\ee
This setup contains no quadratic curvature terms and can be
understood as a very rough approximative analogue of specific
curvature corrected models\footnotetext[9]{The role of
curvature-quartic corrections in M-theory inflation scenarios was
recently discussed in Ref. \cite{MO}.} of M-theory (see e.g.
\cite{EKOY,tseytlin,howe}). The investigation will be performed for
an arbitrary number of dimensions, D.

We start by deriving the explicit form of the potential $U(\phi)$.
For this purpose we substitute $f(\bar R)$ from \rf{4.1} into
relation \rf{1.7},
\be{4.2} f'=e^{A\phi}=1 + 4\gamma \bar R^3,
\ee
and resolve the latter equation for $\bar R$:
\be{4.3} \bar R = (4\gamma)^{-1/3}\left( e^{A\phi}
-1\right)^{1/3},\qquad -\infty < \phi < \infty\; .
\ee
The potential $U(\phi)$ is then found from Eq. \rf{1.8} as
\be{4.5} U(\phi ) = \frac{1}{2}e^{-B\phi} \left[
\frac{3}{4} (4\gamma)^{-1/3}\left(e^{A\phi}-1 \right)^{4/3} +
2\Lambda_D\right]\, .
\ee
{}From its form with
\be{4.5-1}
U[\phi;\Lambda_D>0,\gamma>0 ]\ge 0\quad \forall \phi
\ee
we immediately conclude that the minimum condition $U(\phi_0)<0$
cannot be satisfied in the sector $(\Lambda_D>0,\gamma>0)$. In the
remaining sectors, the potential will have a minimum if the
extremum condition \rf{1.11a} and the minimum-ensuring inequality
\rf{1.11c} will be fulfilled simultaneously. In the present case,
these conditions read
\be{4.4} h[\Lambda_D,\gamma,\bar R]:=\left[\gamma
(D-8)\bar R^4 + (D-2)\bar R -2D\Lambda_D \right]_{\phi_0}= 0
\ee
and
\be{4.4-1}
f''\partial_{\bar R}h=12\gamma \bar R^2\left[(D-2)+4(D-8)\gamma
\bar R^{3}\right]_{\phi_{0}}>0,
\ee
respectively. {}From their structure it follows that the dimension
$D=8$ will constitute an exceptional class of models [due to
cancellation of the highest order terms in \rf{4.4}, \rf{4.4-1}].
We will analyze this class of models in  subsection \ref{d8}
below.

\subsection{Dimensions $D\neq 8$\label{dn8}}

Eq. \rf{4.4}, $h[\Lambda_D,\gamma,\bar R]=0$, is an algebraic
equation in the variables $(\Lambda_D,\gamma,\bar R)$ which
defines a two-dimensional algebraic variety ${\cal V}\subset{\cal
M}$ in the three-dimensional parameter space ${\cal M}=\RR^3\ni
(\Lambda_D, \gamma,\bar R)$. On this variety inequality \rf{4.4-1}
together with the restrictions \rf{1.13-1} and \rf{1.7}
\be{4.4-2}
\bar R<0,\qquad f'=1 + 4\gamma \bar R^3>0
\ee
selects the parameter subset $ \Upsilon \subset {\cal V}$ of
stably compactified internal space configurations. Choosing
$\Lambda_D$ and $\gamma$ as independent parameters, our main task
will consist in obtaining the projection
$\Theta_{(\Lambda_D,\gamma)}:=\pi \Upsilon$ of the stability
region $\Upsilon\subset {\cal V}\subset {\cal M}$ onto the
$(\Lambda_D,\gamma)-$plane. (By $\pi$ we denote the projection
itself.) Most of the information will be derived by finding
restrictions on $(\Lambda_D,\gamma)$ from the conditions which
ensure the reality of $\bar R$ as solution of the algebraic
equation \rf{4.4}.

In order to obtain the solutions $\bar R$ of equation \rf{4.4}
explicitly, we follow standard techniques (see, e.g.,
\cite{Abramowitz} and \ref{app1}) and consider first the associated
cubic equation
\be{4.8} u^3 + \frac{8D\Lambda_D}{\gamma (D-8)}u -
\left[\frac{D-2}{\gamma (D-8)}\right]^2 =0
\ee
and its discriminant $Q$:
\be{4.9}
Q= r^2+q^3,\qquad q := \frac{8D\Lambda_D}{3\gamma (D-8)}, \qquad r
:= \frac{1}{2} \left[\frac{D-2}{\gamma (D-8)}\right]^2\; .
\ee
Depending on the sign of $Q$, the cubic equation \rf{4.8} has one
real solution for $Q>0$ or three real solutions for $Q\le 0$,
where in the case $Q=0$ at least two of these solutions coincide.
Denoting (one of) the real solution(s) by $u_1$, the four roots of
the quartic equation \rf{4.4} can then be obtained according to
\rf{q11}, \rf{q12} as solutions of the two quadratic equations
\be{4.12} \bar R^2 \pm \sqrt{u_1}\; \bar R +\frac 12 \left(
u_1 \pm \epsilon\sqrt{u_1^2 +3q} \right)=0
\ee
with
\be{4.12-2} \epsilon=-\sign\left(\frac{D-2}{\gamma(D-8)}\right).
\ee
Physically sensible solutions will correspond to real roots of
these equations.

Following this general scheme of analysis, we start from the
discriminant $Q$ which we rewrite for later convenience as
\ba{q8}
Q&=&r^2(1+z),\qquad z=z(\Lambda_D,\gamma):=q^3/r^2=4\gamma
(8\Lambda_D/3)^3 w(D),\nn\\ w(D)&:=&\frac{D^3(D-8)}{(D-2)^4}\, .
\ea
In \ref{discrim} it is show that the minimum ensuring inequality
\rf{4.4-1} implies
\be{q8-9n}
z(\Lambda_D,\gamma )> -1
\ee
so that $Q$ is necessarily positive definite, $Q>0$, and, hence,
Eq. \rf{4.8} has only one real-valued root \cite{Abramowitz}
\ba{4.11} u_1 &=& \left[r+Q^{1/2}\right]^{1/3}+
\left[r-Q^{1/2}\right]^{1/3} \nn\\
&=&r^{1/3}v_1(z)>0\, ,
\ea
where
\be{4.11-0}
v_1(z):=\left[1+(1+z)^{1/2}\right]^{1/3}+
\left[1-(1+z)^{1/2}\right]^{1/3}.
\ee
It is now an easy task to explicitly analyze the pair of quadratic
equations \rf{4.12}. They will only have real-valued roots, if at
least one of the corresponding discriminants $\Delta_\pm$ will be
non-negative
\be{4.11-1}
\Delta_\pm =-u_1 \mp 2\epsilon\sqrt{u_1^2+3q}\ge 0\, .
\ee
(The subscripts $\pm$ in $\Delta_\pm$ correspond to the signs in
Eq. \rf{4.12}.) Because of $u_1>0$ this holds only for
\be{4.11-1a}\Delta_{-\epsilon}=-u_1+2\sqrt{u_1^2+3q}
\ee
and under the additional reality-ensuring requirement
\be{4.11-2}
u_1^2+3q\ge 0\, .
\ee
Using the definitions \rf{q8}, \rf{4.11-0} of $z$ and $v_1(z)$,
the inequalities $\Delta_{-\epsilon}\ge 0$ and \rf{4.11-2} can be
reshaped into the form
\ba{4.11-4}
H_1(z)&:=&v_1^6(z)+64z\ge 0\label{4.11-4a}\\
H_2(z)&:=&v_1^6(z)+27z\ge 0\label{4.11-4b}\, ,
\ea
respectively.
\begin{figure}[hbt]
\centerline{\epsfxsize=7cm \epsfbox{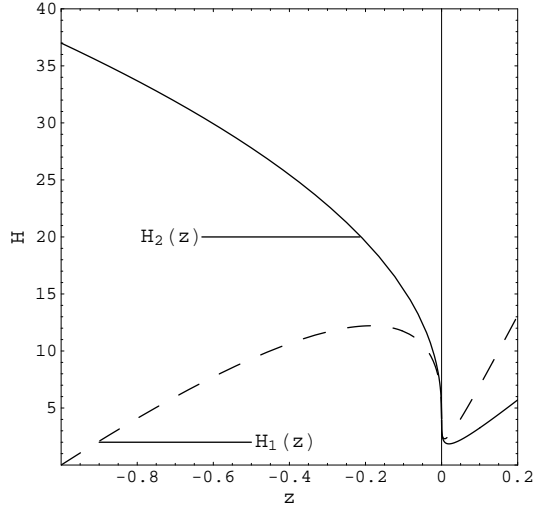} } \caption{Auxiliary
functions $H_1(z)$ and $H_2(z)$, which for $H_{1,2}(z)\ge 0$
ensure a non-negative discriminant $\Delta_{-\epsilon}\ge 0$ of
the quadratic equations \rf{4.12}.\label{fig1}}
\end{figure}

{}From the graphics of the functions $H_{1,2}(z)$ depicted in Fig.
\ref{fig1}, we read off that both inequalities \rf{4.11-4a},
\rf{4.11-4b} are satisfied for $z\ge -1$ and no additional
restrictions are set by them on the allowed region of the
parameter $z$.

Hence, we arrive at the result that the real-valued roots of the
quartic equation \rf{4.4} follow from the equations \rf{4.12} by
identifying in them $\pm=-\epsilon$. The corresponding quadratic
equation reads
\be{4.11-5a}
\bar R^2-\epsilon \sqrt{u_1}\bar R +\frac 12
\left(u_1-\sqrt{u^2_1+3q}\right)=0
\ee
and has solutions (we distinguish them again by subscripts $\pm$)
\ba{4.11-6}
\bar R_{\epsilon,\pm}&=&\frac 12
\left(\epsilon\sqrt{u_1}\pm\sqrt{2
\sqrt{u_1^2+3q}-u_1}\right)\nn\\
&=&r^{1/6}T_{\epsilon,\pm} (z)\, ,
\ea
where
\be{4.11-6a}
T_{\epsilon,\pm} (z):=\frac 12 \left(\epsilon\sqrt{v_1}\pm\sqrt{2
\sqrt{v_1^2+3z^{1/3}}-v_1}\right)\, .
\ee
These roots are defined over the complete parameter region
$z(\Lambda_D,\gamma)>-1$ so that further restrictions on
$(\Lambda_D,\gamma)$ can only follow from the additional
requirements \rf{4.4-2}.

The first of these requirements, $\bar R<0$, should be fulfilled
for a successful freezing stabilization of the extra-dimensional
factor spaces. From the structure of \rf{4.11-6}, \rf{4.11-6a} we
read off that  $\bar R_{+,+}$ contradicts this bound, whereas for
the remaining solutions it partially narrows the allowed parameter
region as follows
\ba{4.11-6b}
\bar R_{+,-}: &\qquad & 0<z,\nn\\
\bar R_{-,+}: &\qquad & -1<z<0,\nn\\
\bar R_{-,-}: &\qquad & -1<z.
\ea

The second inequality, $f'=1 + 4\gamma \bar R^3>0$, of \rf{4.4-2} is
analyzed in \ref{phys-branch} and maps into the following parameter
restrictions
\ba{4.11-15}
D<8:& \qquad &
z<-w(D)=|w(D)|\label{4.11-15a}\, ,\\
D>8:& \qquad & -w(D)<z\label{4.11-15b}\, .
\ea

So far, we have performed our analysis mainly in terms of the
function $z(\Lambda_D,\gamma)$ and, hence, in terms of projections
of the bounds \rf{4.4-1} and $f'>0$ on the
$(\Lambda_D,\gamma)-$plane. For completeness, we have to test
whether all of the projected segments of ${\cal V}$ over the allowed
region of the $(\Lambda_D,\gamma)-$plane fulfill the additional
bound\footnotetext[10]{The restrictions \rf{4.11-6b} are only
necessary conditions for the existence of  solutions $\bar R<0$ of
the quartic equation, and provide only partial information about
\rf{4.4-1} and $f'>0$.} $\bar R<0$, i.e. we should re-analyze
\rf{4.4-1} and $f'>0$ in terms of $\bar R$. We start with
\rf{4.4-1}. A simple case analysis gives
\ba{4.11-16}
\gamma >0:&\qquad & D>8, \ \ \epsilon=-:\qquad \bar
R^3>-\left|\frac{1}{4\gamma}\frac{D-2}{D-8}\right|\label{4.11-16a}\, ,\\
&& D<8, \ \ \epsilon=+:\qquad \bar
R^3<\left|\frac{1}{4\gamma}\frac{D-2}{D-8}\right|\label{4.11-16b}\, ,\\
\nn\\
\gamma <0:&\qquad & D>8, \ \ \epsilon=+:\qquad \bar
R^3>\left|\frac{1}{4\gamma}\frac{D-2}{D-8}\right|>0\label{4.11-16c}\, ,\\
&& D<8, \ \ \epsilon=-:\qquad \bar
R^3<-\left|\frac{1}{4\gamma}\frac{D-2}{D-8}\right|\label{4.11-16d}
\ea
and shows that configurations with $D>8,\gamma<0$ violate the
bound $\bar R<0$, whereas \rf{4.11-16b} is weaker than $\bar R<0$.
The remaining two inequalities \rf{4.11-16a}, \rf{4.11-16d} can be
reshaped with the help of \rf{4.11-6a} and
\be{4.11-16-2}
\left|\frac{1}{4\gamma}\frac{D-2}{D-8}\right|=2^{-3/2}r^{1/2}
\ee
(from Eq. \rf{4.9}) as
\be{4.11-16-4}
T_{-,\pm} >-2^{-1/2},\qquad T_{-,\pm} <-2^{-1/2}\, ,
\ee
respectively.
\begin{figure}[hbt]
\centerline{\epsfxsize=7cm \epsfbox{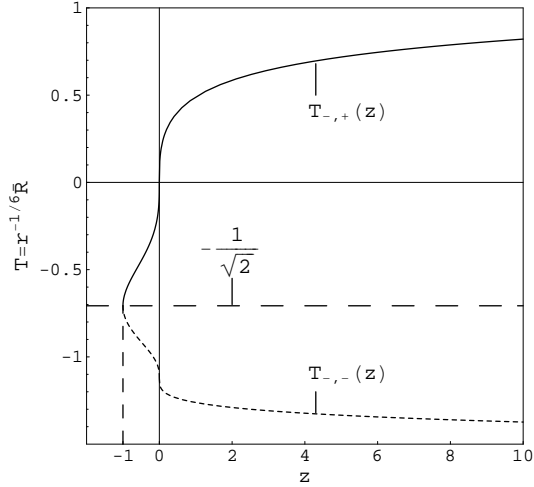} } \caption{Rescaled
scalar curvatures $T_{-,\pm} (z):=r^{-1/6}\bar R_{-,\pm}$.
\label{fig1a}}
\end{figure}
From the graphics of the functions $T_{-,\pm} (z)$ shown in Fig.
\ref{fig1a} we read off that $T_{-,+}(z)>-2^{-1/2}$,
$T_{-,-}(z)<-2^{-1/2}$ for $z>-1$. Hence, inequality \rf{4.11-16a}
is fulfilled only by $\bar R_{-,+}$, whereas \rf{4.11-16d} selects
$\bar R_{-,-}$. A comparison of the inequalities \rf{4.11-16a} -
\rf{4.11-16d} with the condition $f'=1+4\gamma \bar R^3>0$ and its
implications
\be{4.11-17}
\gamma>0:\ -(4\gamma)^{-1}<\bar R^3, \qquad \gamma<0: \ \bar
R^3<|4\gamma|^{-1}
\ee
shows that the latter relations \rf{4.11-17} are compatible with
them and add no additional restrictions.
\begin{figure}[hbt]
\centerline{\epsfxsize=7cm \epsfbox{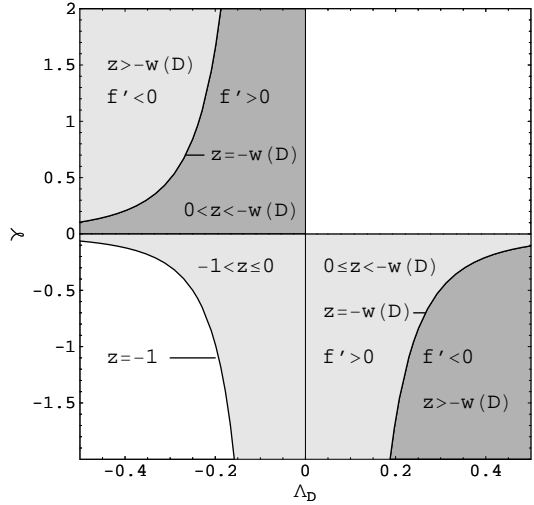} }
\caption{Projection $\Theta_{(\Lambda_D,\gamma)}$ of the stability
region $\Upsilon \subset {\cal V}\subset {\cal M}$ of a
subcritical model with $D<8$ on the $(\Lambda_D,\gamma)-$plane
(shaded areas with $f'>0$). The two lines
$z(\Lambda_D,\gamma)=-w(D)$, given in Eq. \rf{4.11-10}, correspond
to the conformal singularity $f'=0$ where $\bar R \to
-(4\gamma)^{-1/3}$ maps into $R\to-\infty$. They separate
parameter regions with  $U(\phi\to -\infty)\to +\infty$ (dark
grey) from regions with  $U(\phi\to -\infty)\to -\infty$ (light
grey). For $f'>0$ the corresponding regions ensure the existence
of an absolutely stable minimum (dark grey) and a metastable
minimum (light grey, see the discussion in subsection \ref{stab}
below). The $(f'<0)-$regions have been included into the graphics
for completeness.\label{fig2a}}
\end{figure}
\begin{figure}[hbt]
\centerline{\epsfxsize=7cm \epsfbox{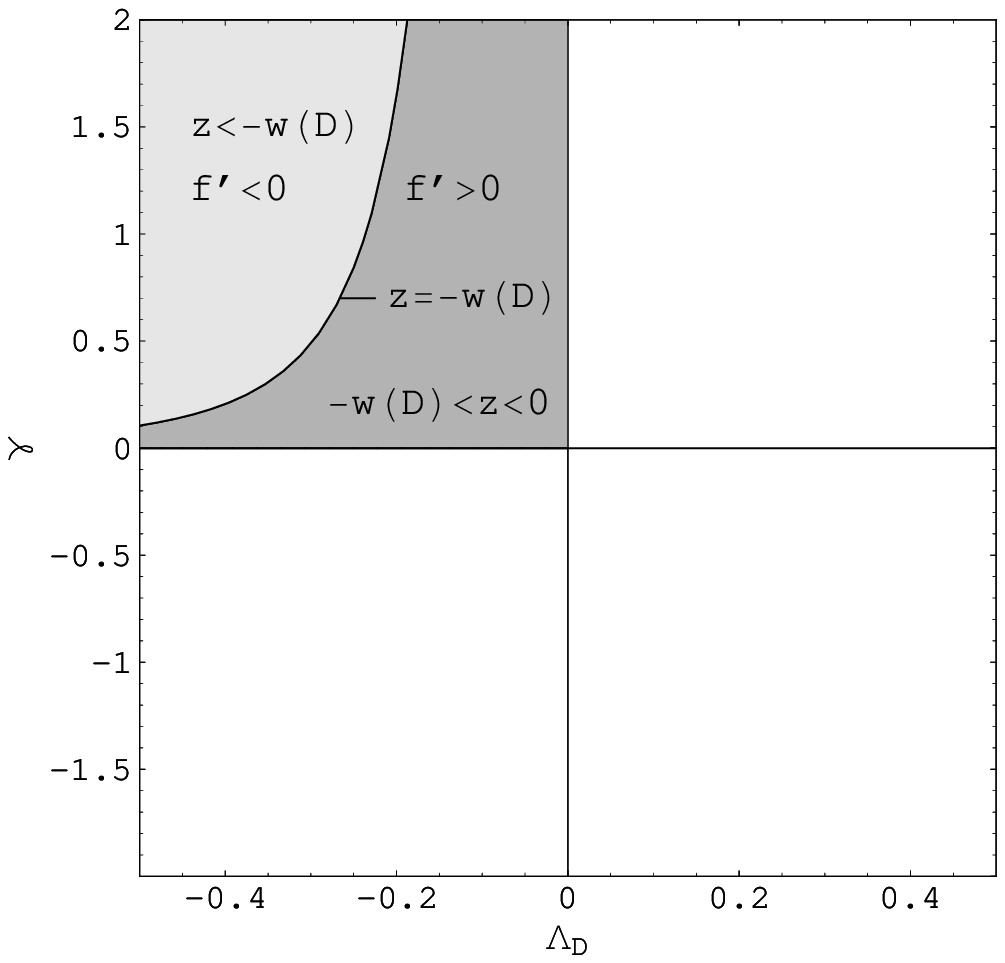} }
\caption{Projection $\Theta_{(\Lambda_D,\gamma)}$ of the stability
region $\Upsilon \subset {\cal V}\subset {\cal M}$ of a model with
$D\ge 8$ on the $(\Lambda_D,\gamma)-$plane (shaded areas with
$f'>0$). The line $z(\Lambda_D,\gamma)=-w(D)$, given in Eq.
\rf{4.11-10}, corresponds to the conformal singularity $f'=0$
where $\bar R \to -(4\gamma)^{-1/3}$ maps into $R\to-\infty$. It
separates the parameter region with  $U(\phi\to -\infty)\to
+\infty$ (dark grey) from the region with  $U(\phi\to -\infty)\to
-\infty$ (light grey). For $f'>0$ the corresponding region ensures
the existence of an absolutely stable minimum (dark grey). The
$(f'<0)-$region is depicted for completeness. The lower half-plane
$\gamma<0$ corresponds to unstable configurations.\label{fig2b}}
\end{figure}

Combining the information about the inequalities \rf{4.11-16a} -
\rf{4.11-16d} with \rf{4.5-1} and the relations \rf{4.11-6b},
\rf{4.11-15a}, \rf{4.11-15b} we arrive at the following parameter
regions for configurations with a possible freezing stabilization
of extra-dimensional (internal) factor spaces:
\ba{4.11-18}
D>8:&\qquad  \gamma >0:\qquad &  \bar
R_{-,+}, \quad -w(D)<z<0\label{4.11-18a}\, ,\\
&\qquad \gamma <0:\qquad & \mbox{no stability}\label{4.11-18b}\, , \\
D<8:&\qquad \gamma>0:\qquad & \bar R_{+,-}, \quad
0<z<|w(D)|\label{4.11-18c}\, ,\\
&\qquad \gamma<0:\qquad & \bar R_{-,-}, \quad
-1<z<|w(D)|\label{4.11-18d}\, .
\ea
This result is schematically shown in Figs.
\ref{fig2a},\ref{fig2b}. (The regions of formal extension to
configurations with $f'<0$ are included in the graphics for
reasons of completeness. They are briefly discussed in subsection
\ref{stab}, below.)

\subsection{The exceptional dimension $D=8$\label{d8}}

In this particular case, the $\bar R-$nonlinear terms in the
minimum-ensuring conditions \rf{4.4}, \rf{4.4-1}
cancel\footnotetext[11]{For completeness we note that in the setup
of the previous subsection \ref{dn8} the formal limit $D\to 8$ (for
$D$ assumed as non-discrete and real-valued) would correspond to the
exceptional (singular) case $z\to 0$, $Q\to +\infty$.} and the
bounds on the solution can be read off immediately:
\be{e1}
\bar R = \frac{8}{3}\Lambda_D<0, \qquad \gamma >0\, .
\ee
[$\bar R<0$ follows from relation \rf{1.12-1} and condition
\rf{1.13-1}.] The restriction $f'=1+4\gamma \bar R^3>0$ sets the
same additional bound on the allowed parameter region
\be{e2}
-1<4\gamma \left(\frac 83 \Lambda_D\right)^3 <0
\ee
as \rf{4.11-15b} in the case of $D>8$.

\subsection{Stable and metastable configurations\label{stab}}

In the previous two subsections \ref{dn8} and \ref{d8}, it has
been shown that models in $D\ge 8$ dimensions posses a stability
region $\Theta_{(\Lambda_D,\gamma)}=\pi \Upsilon$, which is
located in the upper $(\Lambda_D,\gamma)-$half-plane. For models
in $D<8$ dimensions this region is larger and extends additionally
into the lower $(\Lambda_D,\gamma)-$plane (see the relations
\rf{4.11-18a} - \rf{4.11-18d} and Figs. \ref{fig2a}, \ref{fig2b}).
Now we will analyze the system over these stability regions in
more detail.

We start with the asymptotic behavior of the potential $U(\phi)$ in
the limits $\phi\to \pm \infty$. In the high curvature
limit\footnotetext[12]{Here and in the subsequent considerations the
high curvature limits are understood as formal limits (in the
mathematical sense) within the framework of our simplified toy
model. It is clear that in a model with scalar curvature
nonlinearity of the type $f(\bar R)=\bar R + \sum_{k=2}^N a_k \bar
R^k$ gravitational self-interaction effects will dominate when
$\sum_{k=2}^N a_k \bar R^{k-1}\gtrsim 1$. (For discussions of
related subjects we refer to \cite{qg-2}.) This means that a
self-consistent treatment of the model would require techniques from
(loop) quantum gravity or the high-energy sector of M-theory ---
what is out of the scope of the present paper.} $\phi\to +\infty$
with dominant nonlinearity of the type $f'=e^{A\phi}=1+4\gamma \bar
R^3 \gg1$ we find from \rf{4.5}
\be{a3}
U(\phi \to +\infty)\approx \frac 38 (4\gamma)^{-1/3}
e^{(-B+4A/3)\phi}
\ee
and the sign of the coefficient in the exponent
\be{a4}
\frac 43 A-B=\frac A3 \frac{D-8}{D-2}
\ee
leads for fixed $\gamma$ to a qualitatively different behavior for
dimensions $D>8$ and $D<8$:
\be{a5}
U(\phi \to +\infty)\to \frac 38 (4\gamma)^{-1/3}\times\left\{
\begin{array}{c}
  \infty \\
  1 \\
  0
\end{array}\right. \quad \begin{array}{c}
  \mbox{for}\ D>8\, , \\
  \mbox{for}\ D=8\, , \\
  \mbox{for}\ D<8\, .
\end{array}
\ee
The existence of a critical dimension (in our case $D=8$) is a
rather general feature of gravitational theories with polynomial
scalar curvature terms (see, e.g., Refs. \cite{BC,paul}). It can
be easily demonstrated for a model with curvature nonlinearity of
the type
\be{a6}
f(\bar R)=\sum_{k=0}^N a_k \bar R^k
\ee
for which the ansatz
\be{a7}
e^{A\phi}=f'=\sum_{k=0}^N k a_k \bar R^{k-1}
\ee
leads, similar like \rf{1.8a}, to a potential
\be{a8}
U(\phi)=\frac12 \left(f'\right)^{-D/(D-2)}\sum_{k=0}^N (k-1)a_k
\bar R^k.
\ee
In the limit $\phi\to +\infty$ the curvature will behave like
$\bar R\approx c e^{h \phi}$ where $h$ and $c$ can be defined from
the dominant term in \rf{a7}:
\be{a8-1}e^{A\phi}\approx N a_N \bar
R^{N-1}\approx N a_N c^{N-1} e^{(N-1)h\phi}.
\ee
Here the requirement $f'>0$ allows for the following sign
combinations of the coefficients $a_N$ and the curvature
asymptotics $\bar R (\phi\to \infty)$:
\ba{a8-2}
N=2l: & \qquad & \sign [a_N ] =  \sign [\bar R (\phi\to
\infty)]\nn\\
N=2l+1: & & a_N >0, \quad \sign [\bar R (\phi\to \infty)]=\pm 1.
\ea The other combinations, $N=2l: \ \ \sign [a_N ] =  -\sign
[\bar R (\phi\to \infty)]$, \  $N=2l+1: \ \ a_N <0, \ \sign [\bar
R (\phi\to \infty)]=\pm 1$, would necessarily correspond to the
$f'<0$ sector, so that the complete consideration should be
performed  in terms of the extended conformal transformation
technique of Ref. \cite{Maeda}. Such a consideration is out of the
scope of the present paper and we restrict our attention to the
cases \rf{a8-2}. The coefficients $h$ and $c$ are then easily
derived as $h=A/(N-1)$ and $c=\sign(a_N)\left|N
a_N\right|^{-\frac{1}{N-1}}$. Plugging this into \rf{a8} one
obtains
\be{a9}
U(\phi\to +\infty)\approx
\sign(a_N)\frac{(N-1)}{2N}\left|Na_N\right|^{-\frac{1}{N-1}}
e^{-\frac{D}{D-2}A\phi}e^{\frac{N}{N-1}A\phi}
\ee
and that the
exponent \be{a10} \frac{D-2N}{(D-2)(N-1)}A \ee changes its sign at
the critical dimension $D=2N$:
\be{a11}  U(\phi \to +\infty)\to
\sign(a_N)\frac{(N-1)}{2N}\left|Na_N\right|^{-\frac{1}{N-1}}\times
\left\{ \begin{array}{c}
  \infty \\
  1 \\
  0
\end{array}\right. \quad \begin{array}{c}
  \mbox{for}\ D>2N\, , \\
  \mbox{for}\ D=2N\, , \\
  \mbox{for}\ D<2N\, .
\end{array}
\ee This critical dimension
$D=2N$ is independent of the concrete coefficient $a_N$ and is
only defined by the degree $\deg_{\bar R} (f)$ of the scalar
curvature polynomial $f$. {}From the asymptotics \rf{a11} we read
off that in the high curvature limit $\phi\to +\infty$, within our
oversimplified classical framework, the potential $U(\phi)$ of the
considered toy-model shows asymptotical freedom for subcritical
dimensions $D<2N$, a stable behavior for $a_N>0$, $D>2N$ and a
catastrophic instability for $a_N<0$, $D>2N$. We note that this
general behavior suggests a way how to cure a pathological
(catastrophic) behavior of polynomial $\bar R^{N_1}-$nonlinear
theories in a fixed dimension $D>2N_1$: By including higher order
corrections up to order $N_2>D/2$ the theory gets shifted into the
non-pathological sector with asymptotical freedom. More generally,
one is even led to conjecture that the partially pathological
behavior of models in supercritical dimensions could be an
artifact of a polynomial truncation of an (presently unknown)
underlying non-polynomial $f(\bar R)$ structure at high curvatures
--- which probably will find its resolution in a strong coupling
regime of $M-$theory or in loop quantum gravity.
\begin{figure}[hbt]
\centerline{\epsfxsize=7cm \epsfbox{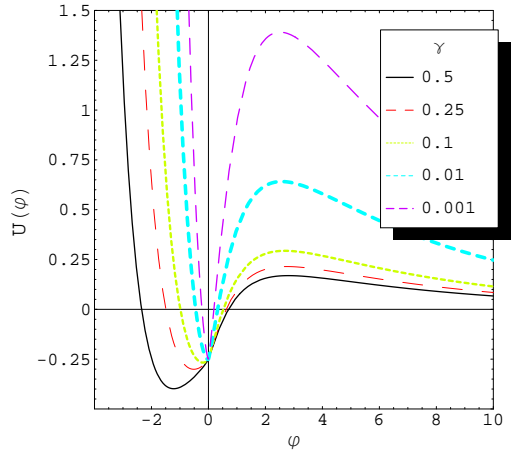} } \caption{Typical
form of a potential $U(\phi)$ with parameters $(\Lambda_D,\gamma)$
from a subcritical $(D<8)$ region of absolute stability
$U(\phi\to-\infty)\to +\infty$, $U(\phi\to +\infty)\to +0$ in the
physical sector $f'>0$. Specifically,  it is set $D=6$,
$\Lambda_D=-1/4$ for several values of $\gamma$ (in Planck
units).\label{fig3}}
\end{figure}
\begin{figure}[hbt]
\centerline{\epsfxsize=7cm \epsfbox{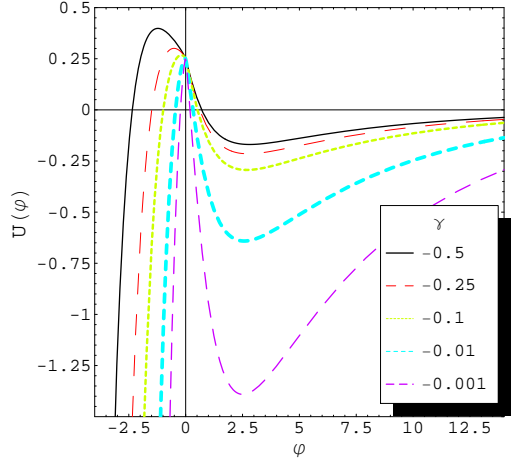} } \caption{Typical
form of a potential $U(\phi)$ of a subcritical $(D<8)$ metastable
system with  $U(\phi\to-\infty)\to -\infty$, $U(\phi\to
+\infty)\to -0$ in the physical sector $f'>0$. It is set $D=6$,
$\Lambda_D=1/4$ for several values of $\gamma$ (in Planck
units).\label{fig4}}
\end{figure}

As next step we consider the opposite limit $\phi\to -\infty$
which corresponds to $f'\to 0$. {}From \rf{1.8a} we see that the
potential $U(\phi)$ of a model with general polynomial $\bar
R-$nonlinearity \rf{a6} behaves in this limit like
\be{a12}
U(\phi\to -\infty)\approx -\frac 12 e^{-\frac{D}{D-2}A\phi}f(\bar
R_{ci})\, ,
\ee
where $\bar R_{ci}$ is one of the real roots of the polynomial
$f'(\bar R)=0$ given in \rf{a7}. In other words, the potential
$U(\phi)$ diverges like
\be{a13}
U(\phi\to -\infty)\to -\sign[f(\bar R_{ci})]\times \infty
\ee
for any dimension $D$ and any value $f(\bar R_{ci})\neq 0$. This
means that, for energetic reasons, the system will be repelled
from configurations with $f'\approx 0$ by an infinitely high
barrier $U(\phi\to -\infty)\to +\infty$ in the case $f(\bar
R_{ci})<0$, and it will be catastrophically attracted (experience
a collapse) to $f'\approx 0$ in the case $f(\bar R_{ci})>0$, \
$U(\phi\to -\infty)\to -\infty$.

Let us make this general consideration explicit for the $\bar
R^4-$model \rf{4.1}. The polynomial $f'=1+4\gamma \bar R^3$ has
the single real-valued root $\bar R_{c2}=-(4\gamma)^{-1/3}$, which
was used in \rf{4.11-7} - \rf{4.11-15b} to map the inequality
$f'>0$ into the $(\Lambda_D,\gamma)-$plane [result: the bounds
$-w(D>8)<z$, \ $z<-w(D<8)$]. Plugging this root into $f$ we get
\be{a14}
f(\bar R_{c2})=-\left[\frac 34 (4\gamma)^{-1/3}+2\Lambda_D\right].
\ee
{}From the condition $f(\bar R_{c2})<0$ for the existence of a
repelling potential barrier with $U(\phi\to -\infty)\to +\infty$
limit one finds the following inequalities (regardless of the
existence of a minimum)
\ba{a15}
\gamma>0:&\qquad & \frac 83
(4\gamma)^{1/3}\Lambda_D>-1\, ,\nn\\
\gamma<0:&\qquad & \frac 83 (4\gamma)^{1/3}\Lambda_D<-1\, .
\ea
(Below we will show that the case $\gamma>0$ holds for the
$(f'>0)-$sector, whereas $\gamma<0$ will correspond to $f'<0$.)
Multiplication with $w(D)$ leads to the equivalent conditions
\ba{a16}
D>8:&\qquad  \gamma >0:\qquad &   -w(D)<z\label{a16-a}\, ,\\
&\qquad \gamma <0:\qquad & z<-w(D)\label{a16-b}\, , \\
D<8:&\qquad \gamma>0:\qquad & z<-w(D)=|w(D)|\label{a16-c}\, ,\\
&\qquad \gamma<0:\qquad & |w(D)|=-w(D)<z\label{a16-d}\, .
\ea
We see that the asymptotical behavior of the potential $U(\phi\to
-\infty)$ is defined by similar parameter regions in $z$ like those
which control the existence of a minimum under the condition $f'>0$.
In both cases the bound is connected with the critical value $f'=0$
which corresponds to $z_c=-w(D)$. This is not a surprise because the
two simultaneous conditions $f'(\bar R_{c2})=0$, $f(\bar R_{c2})=0$,
which define the critical value $z_c$ in the inequalities $f(\bar
R_{c2})<0$,  $f(\bar R_{c2})>0$, fulfill the extremum condition
$\partial_\phi U(\phi)=0$, i.e. $h=Df-2f'\bar R=0$ in \rf{1.11a} and
hence the quartic equation\footnotetext[13]{Obviously, it holds
trivially $h=Df-2f'\bar R=0$ for $\bar R=\bar R_{c2}$.} \rf{4.4}.

Analogously to inequalities \rf{a15} - \rf{a16-d} we get from the
condition $f(\bar R_{c2})>0$ for a catastrophically attracting
potential, $U(\phi\to -\infty)\to -\infty$, that such an
asymptotics holds over the sectors
\ba{a17}
D>8:&\qquad  \gamma >0:\qquad &   z<-w(D)\label{a17-a}\, ,\\
&\qquad \gamma <0:\qquad & -w(D)<z\label{a17-b}\, , \\
D<8:&\qquad \gamma>0:\qquad & |w(D)|=-w(D)<z\label{a17-c}\, ,\\
&\qquad \gamma<0:\qquad & z<-w(D)=|w(D)|\label{a17-d}\, .
\ea
These sectors are complementary to those in \rf{a16-a} - \rf{a16-d}.
Here, we observe that  $\gamma<0$ for $f'>0$ and $\gamma>0$ for
$f'<0$. This is confirmed by a comparison of inequalities \rf{a16-a}
- \rf{a17-d} with \rf{4.11-15a}, \rf{4.11-15b} and the inequalities
in \ref{f-}. Obviously, the regions $z<-w(D)$ for $D>8$ and
$z>-w(D)=|w(D)|$ for $D<8$ correspond to a formal extension of the
potential $U(\phi)$ into the $(f'<0)-$sector. For completeness,
these regions have been included into Figs. \ref{fig2a},
\ref{fig2b}. The typical $U(\phi\to -\infty)\to \pm \infty$ behavior
of the potential is illustrated in Figs. \ref{fig3},\ref{fig4}.

{}From Fig. \ref{fig4} we see that the minimum is separated by a
barrier of finite height and width from the singularity at
$\phi\to -\infty$. Hence, we find that the $\bar R^4-$models in
the $(f'>0)-$sector are absolutely stable for $\gamma>0$ and
metastable with tendency to collapse into the singularity
$f'\approx 0$ for $\gamma<0$. We further see from the conformal
relation \rf{1.10} between the scalar curvature $\bar R$ of the
nonlinear model and the curvature $R$ of the equivalent linear
model that $f'=0$, and hence $z=|w(D)|$, corresponds to a
conformal singularity: The finite curvature value $\bar
R_{c2}=-(4\gamma)^{-1/3}$ of the nonlinear model is related to a
curvature singularity $R\sim (f')^{-2/(D-2)}\bar R_{c2}$ in the
associated linear model.

A detailed study of various limiting cases is given in \ref{limits}.
The corresponding main results can be summarized as follows.
\begin{itemize}
\item The limit $(\Lambda_D\to -0,\gamma>0)$ in the stable sector corresponds
to a flat-space limit $\bar R\to -0$ which via \rf{4.2} is
associated with a freezing of the nonlinearity field $\phi$:\ \
$f'\to 1$ at $\phi_0\to 0$. In the metastable sector nothing special
happens in the limit $(\Lambda_D\to 0,\gamma<0)$.
\item For $\Lambda_D\neq 0$ the limit $\gamma\to +0$ corresponds to
a freezing of the nonlinearity field $\phi$ at $\phi_0=0$ and a
smooth transition to a linear gravity model of Einstein-Hilbert
type. In contrast, the limit $\gamma\to -0$ of the metastable
$(D<8)-$system results in an infinitely deep minimum
$U(\phi_0,\gamma\to -0)\to -\infty$ at $\phi_0=(1/A) \ln|3D/(D-8)|$
and a curvature singularity $\bar R_{-,-}(\gamma\to -0,D<8)\approx
-|(D-2)/[\gamma(D-8)]|^{1/3}$.
\item The limit $z\to -1$ (in the metastable sector) corresponds
to coalescing minimum and maximum of the potential $U(\phi)$ (with
resulting inflection point at $z=-1$). For $z\le -1$ the potential
$U(\phi)$ has no minimum at all and the system is completely
unstable.
\end{itemize}

\subsection{Inflation in the {\boldmath $(\gamma>0)-$}sector\label{inflation}}

\begin{figure}[tbh]
\centerline{\epsfxsize=7cm \epsfbox{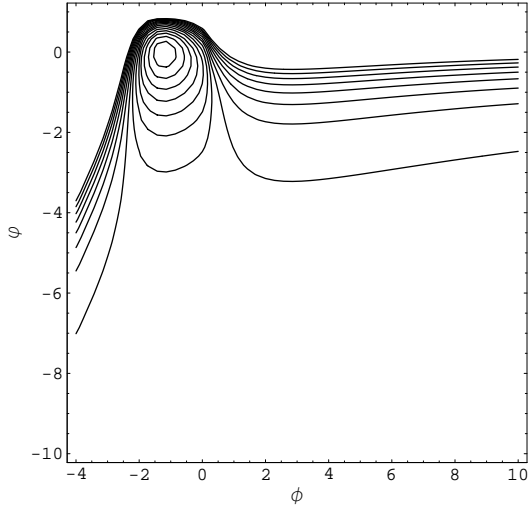} } \caption{Contour
plot of the effective potential $U_{eff}(\phi,\varphi)$ given in
Eq. \rf{4b.8-1}.\label{fig5}}
\end{figure}

\begin{figure}[bth]
\centerline{\epsfxsize=7cm \epsfbox{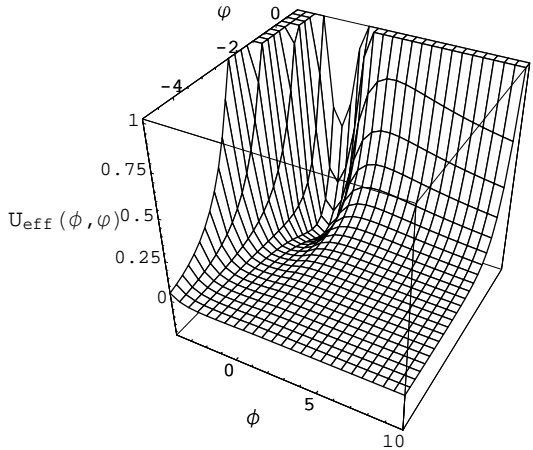} } \caption{3D plot
of the effective potential $U_{eff}(\phi,\varphi)$ given in
Eq.\rf{4b.8-1}.\label{fig6}}
\end{figure}

For simplicity we restrict our attention to most  simple models
with only one internal factor space $M_1$. The requirement for
non-vanishing negative curvature of this space (in order to ensure
a late-time stabilization of the corresponding dimensions) holds
only for total dimensions $D\ge 6$ (in the case of $D_0=4$). In
terms of the normalized scale factor (radion) of this internal
space $M_1$
\be{4b.8}
\varphi \equiv - \sqrt{\frac{d_1(D-2)}{D_0-2}}\hat \beta^1
\ee
the effective potential \rf{1.31} reads
\ba{4b.8-1}
U_{eff}=e^{2s\varphi}\left[U(\phi)-\frac 12 \hat R_1 e^{2s_1\varphi}
\right], &\qquad & s:=\sqrt{\frac{d_1}{(D_0-2)(D-2)}}, \nn\\ &&
s_1=\frac{D_0-2}{d_1}s\, .
\ea
In Figs. \ref{fig5},  \ref{fig6} its generic form is illustrated
by a model with $(d_1=2)$ extra dimensions and parameters $\gamma
=1/2$, $\Lambda_D=-1/4$.
A general feature of an effective potential \rf{4b.8-1} with a
minimum  and a barrier asymptotic $U(\phi\to -\infty)\to +\infty$
is the necessary absence of a local maximum or a saddle point.
This is easily seen from the extremum conditions
\ba{6.1}
\left.\partial_\varphi U_{eff}\right|_{extr}&=&\left.2s
e^{2s\varphi}\left[U(\phi)+\frac{D_0-2-d_1}{2d_1}\hat R_1
e^{2s_1\varphi}\right]\right|_{extr}=0\, ,\label{6.1-a}\\
\left.\partial_\phi
U_{eff}\right|_{extr}&=&e^{2s\varphi}\left.\partial_\phi
U(\phi)\right|_{extr}=0\label{6.1-b}
\ea
and their implications.  Condition \rf{6.1-b} yields again the
quartic equation \rf{4.4}. For $D>8$, $\gamma>0$ this equation has
only a single solution $\bar R_{-,+}$ in the
$(f'>0)-$sector\footnotetext[14]{From Fig. \ref{fig1a} and relation
\rf{4.11-18a} one sees that the condition $f'>0$ cuts the
$T_{-,+}(z)-$branch at $z_{c2}=-w(D)$ (where $-1<z_{c2}<0$) and
allows only for the piece $-w(D)<z<0$ where
$T_{-,+}(z)>T_{-,+}(z_{c2})$ holds. The remaining segment
$-1<z<-w(D)$ with $T_{-,+}(z)<T_{-,+}(z_{c2})$ and also the whole
second solution $T_{-,-}(z)$ are completely located in the
$(f'<0)-$sector.} so that the  minimum is the only extremum. In the
case $D<8$, $\gamma>0$ beside the negative curvature solution $\bar
R_{+,-}$ of the minimum there exists a positive maximum $\bar
R_{+,+}> 0$ (see Eq. \rf{4.11-6}) so that at the corresponding
extremum with regard to $\phi$ it holds $U(\phi_{max})>0$ in
accordance with Eq. \rf{1.12-1}. Hence, the condition \rf{6.1-a}
cannot be fulfilled due to $U(\phi_{max})>0$, $\hat R_1<0, \  D_0=4,
\ d_1\ge 2 \  \ \Longrightarrow \ \
\partial_\varphi U_{eff}>0$ and there will be no other
extremum of the effective potential $U_{eff}(\phi,\varphi)$ apart
from the already studied minimum at $(\phi=\phi_0,\varphi=0)$.

This means that in the considered oversimplified model inflation
of purely topological type \cite{topinfl} as, e.g., recently
demonstrated for SUGRA inspired setups (racetrack inflation
starting at a saddle point of the effective potential) in Ref.
\cite{racetrack1} is ruled out. In general, the too steep slopes
of the exponential terms of the effective potential will spoil
inflation, i.e. a slow-roll behavior seems not realistic. We will
demonstrate this with a region in the $(\phi,\varphi)-$plane where
one still might hope to obtain a sufficiently gentle slope to
induce the needed long-lasting accelerated expansion of the
external space as well as an attraction to the global minimum (in
order to ensure a late-time stabilization of the scalar fields).
(See Figs. \ref{fig5}, \ref{fig6}.) Such a region  could be
expected close to the maximum of the potential $U(\phi)$ where in
rough approximation holds
\be{4b.9}
U_{eff} \approx e^{2s\varphi} U(\phi )\, .
\ee
The action functional \rf{1.27} shows that the two scalar fields
$\varphi_1:=\phi,\varphi_2:=\varphi$ live in a flat
($\sigma-$model) target space. Hence, the estimate of the slow
roll parameters can be performed within a simplified version of
the multi-field inflation scheme of Refs.
\cite{multi-1,multi-2,multi-inflation}. Assuming  in rough
approximation that the external space has already flattened, the
inflation parameters $\epsilon, \ \eta$ read
\ba{4b.10}
\epsilon =-\frac{\dot H}{H^2} &\approx & \frac 12 \frac{|\partial
U_{eff}|^2}{U_{eff}^2}\label{4b.10a}\\
\eta =-\frac{\sum_{i=1}^2\ddot \varphi_i\dot\varphi_i}{H|\dot
\varphi|^2}&\approx &
-\epsilon+\frac{\sum_{i,j=1}^2(\partial_{ij}^2 U_{eff})(\partial_i
U_{eff})(\partial_j U_{eff})}{U_{eff} |\partial U_{eff}
|^2}\label{4b.10b}.
\ea
Here, $H=\dot a/a$ is as usual the Hubble parameter of the
external space.  Additionally, the following abbreviations have
been introduced
\be{4b.11}
\partial_i:=\frac{\partial}{\partial \varphi_i},\qquad |\partial
U_{eff}|^2=\sum_{i=1}^2\left(\partial_i U_{eff}\right)^2,\qquad
|\dot \varphi|^2=\sum_{i=1}^2 \dot\varphi_i^2\, .
\ee
Inflation is possible for $\epsilon<1$ and a sufficiently small
slow-roll parameter $|\eta|\ll 1$. In the vicinity of the maximum
in $\phi-$direction it holds $\partial_\phi U|_{max}\approx0$ so
that $\epsilon,\ \eta$ are essentially defined by the slope in
$\varphi-$direction. Explicitly one obtains in this region
\be{4b.12}
\epsilon \approx  2s^2,\qquad  \eta \approx  2s(1-s).
\ee
For a four-dimensional external space $D_0=4$ this yields from
\rf{4b.8-1}
\be{4b.12-2}
s^2=\frac{d_1}{2(d_1+2)}<\frac 12
\ee
and the rough estimates
\ba{4b.12-3}
\begin{array}{rll}
  d_1=2:\qquad & \epsilon \approx 0.5\, ,\quad & \eta \approx 0.5\, ,
  \\[.5ex]
  d_1=3:\qquad & \epsilon \approx 0.6\, ,\quad & \eta \approx 0.49\, ,
  \\[.5ex]
  d_1=6:\qquad & \epsilon \approx 0.75\, ,\quad & \eta \approx 0.47\ . \\
\end{array}
\ea
Because $|\eta|\ll 1$ is not satisfied, the considered toy model
would produce an accelerated expansion $(\epsilon<1)$ which would
be much too short for successful inflation. Whether a domain wall
with possibility for topological type inflation could form between
regions to the left and to the right (in $\phi-$direction) of the
crest at the maximum of $U(\phi)$, remains an open question.
Corresponding indications have been given in Ref. \cite{EKOY}, but
seem to require an additional detailed analysis --- and probably
an embedding of the toy model into a more general setup with
richer structure.

\section{Conclusion\label{conclu}}

In the present paper we continued our investigation
\cite{GMZ1,GMZ2} on multidimensional gravitational models with a
non-Einsteinian form of the action. The corresponding action
functional was assumed as a smooth function $f(\bar R)$ of the
scalar curvature $\bar R$ of a $D-$dimensional spacetime manifold
with warped product structure. The main subject of our
considerations was the stabilization problem for the extra
dimensions. As technique we used a reduction of the nonlinear
gravitational model to a linear one with an additional
self-interacting scalar field (nonlinearity scalar field $\phi$).
The factorized geometry allowed for a dimensional reduction of the
considered model and a transition to the Einstein frame. As
result, we obtained an effective four--dimensional model with
nonlinearity scalar field and additional minimally coupled scalar
fields which describe conformal excitations of the scale factors
of the internal space (its zero-mode excitations).

In terms of these scalar fields we performed a detailed stability
analysis for models with scalar curvature nonlinearities of the
type $f(\bar R)=\bar R-\mu /\bar R$, $\mu>0$ and $f(\bar R)=\bar
R+\gamma \bar R^4-2\Lambda_D$, where $\Lambda_D$ plays the role of
a $D-$dimensional bare (bulk) cosmological constant. As stability
condition we assumed the existence of a minimum of the effective
potential of the dimensionally reduced theory so that a late-time
attractor of the system could be expected with freezing
stabilization of the extra-dimensional scale factors and the
nonlinearity field. It was shown in Refs. \cite{GMZ1,GMZ2}, that
for purely geometrical setups this is only possible for negative
scalar curvatures, $(\bar R<0)$, independently of the concrete
form of the function $f(\bar R)$.

Four-dimensional purely gravitational models with $\bar R^{-1}$
curvature contributions have been proposed recently as possible
explanation of the observed late-time acceleration (dark energy)
of the Universe \cite{1/R-1}. In section \ref{first model} of the
present paper, we showed that higher dimensional models with the
same $\bar R^{-1}$ scalar curvature nonlinearity reproduce (after
dimensional reduction) the two solution branches of the
four-dimensional models. But due to their oversimplified structure
these models cannot simultaneously provide a late-time
acceleration of the external four-dimensional spacetime and a
stabilization of the internal space. A late-time acceleration is
only possible for one of the solution branches
--- for that which yields  a positive maximum of the potential
$U(\phi)$ of the nonlinearity field. A stabilization of the
internal spaces requires a negative minimum of $U(\phi)$ as it can
be induced by the other solution branch. The question of whether
this incompatibility could be resolved by different scalar
curvature nonlinearities over the factor spaces (for each factor
space $M_i$ it might hold its own curvature nonlinearity $f_i(\bar
R_i)$) is still open and deserves a separate analysis. We left
this issue to future investigations.

The considered $\bar R^4-$setup was assumed as highly simplified toy
model analogue of the loop corrected gravity sector of $M-$theory
\cite{howe}. The stability analysis of the higher dimensional model
was reduced to a set of algebraic compatibility tests for the
extremum condition (in the present case a quartic equation in the
scalar curvature $\bar R$) and inequalities which ensure the
existence of a minimum of the effective potential (non-tachyonic
mass terms of the corresponding field excitations). For simplicity,
we restricted the investigation to parameter regions $f'=1+4\gamma
\bar R^3>0$ which are smoothly connected with the curvature-linear
model at $f'=1$ without passing a conformal singularity. In
Brans-Dicke (Jordan) frame the latter requirement ensures an
effective gravitational constant which is positive
definite\footnotetext[14]{See footnote \ref{f-} for a few comments
on the BD-antigravity sector with $f'<0$.} and smoothly connected
with that of a given BD frame model with a fixed (frozen)
gravitational constant.

With the help of a projection technique in the
$(\Lambda_D,\gamma,\bar R)-$space ${\cal M}$ we identified regions
which ensure the existence of a minimum of the effective
potential, and hence fulfill a necessary condition for a
successful freezing stabilization of the extra dimensions. The
results can be summarized as follows. For systems with total
spacetime dimensions $D\ge 6$ (in the case of a
$(D_0=4)-$dimensional external spacetime) there exists a stable
sector $\Theta_{1,(\Lambda_D,\gamma)}=\left\{\Lambda_D<0\ \cap \
\gamma>0 \ \cap \ |z(\Lambda_D,\gamma)|<|w(D)|\right\}$ on the
$(\Lambda_D,\gamma)-$plane which in the limit $\gamma\to +0$ tends
smoothly to the $\bar R-$linear sector. The corresponding
transition is connected with a freezing of the nonlinearity field
at the minimum of its potential $U(\phi)$, i.e. a diverging
excitation mass, $m_\phi^2\to +\infty$, due to a diverging Hessian
of the potential $U(\phi)$. Models within
$\Theta_{1,(\Lambda_D,\gamma)}$ are separated from the conformal
singularity at $f'=0$ (and the antigravity sector $f'<0$ beyond
it) by a potential barrier of infinite height and width and are,
hence, absolutely stable with regard to transitions into the
$(f'<0)-$sector. Additionally, it was shown that  the limit
$\Lambda_D\to -0$ in the $\Theta_{1,(\Lambda_D,\gamma)}-$sector,
is connected with a decompactification of the internal space
components $M_i, \ i=1,\ldots, n$ and a flattening $\bar R\to -0$
of the bulk spacetime $M$.

Apart from $\Theta_{1,(\Lambda_D,\gamma)}$ there exists a second
stability sector $\Theta_{2,(\Lambda_D,\gamma)}=\left\{\gamma<0 \
\cap \ -1<z(\Lambda_D,\gamma)<-w(D)\right\}$ for dimensions $D<8$.
The potential $U(\phi)$ for such configurations is unbounded from
below in the limit $f'(\phi\to -\infty)\to +0$ and has a minimum
which is separated from the conformal singularity at $f'=0$ (and
the antigravity sector $f'<0$ beyond it) by a potential wall of
finite height and width. Configurations in this minimum would be
metastable and  prone to collapse into $f'=0$. The
$\Theta_{2,(\Lambda_D,\gamma)}-$sector is disconnected from
$\Theta_{1,(\Lambda_D,\gamma)}$ by an essential singularity of
$\bar R$ and $U(\phi_0)$ in the limit $\gamma\to -0$. In this
limit the nonlinearity field $\phi$ freezes at $\phi_0(\gamma\to
-0)\to (1/A)\ln(3D/|D-8|)$ with $m^2_\phi \to +\infty$, but
simultaneously the scalar curvature diverges $\bar R\to -\infty$
and the potential deepens unboundedly $U(\phi_0)\to -\infty$. This
behavior is a strong indication for inconsistencies of
$\Theta_{2,(\Lambda_D,\gamma)}-$configurations within the
framework of the given limited setup. The question of whether the
$\Theta_{2,(\Lambda_D,\gamma)}-$sector of the considered
oversimplified toy-model will find a physically sensible
interpretation within a still unknown extended curvature-nonlinear
theory of gravity, a special UV limit of non-perturbative M-theory
or within loop quantum gravity remains an open issue.

A further issue which was out of the scope of the present paper was
the analysis of dynamical transitions between configurations which
correspond to different solution branches $\bar R (\phi)$ of Eq.
\rf{1.7}, $f'(\bar R)=e^{A\phi}$. For the $\bar R^{-1}-$model of
section \ref{first model}, e.g., there exist two such branches $\bar
R(\phi)$ which form a double cover\footnotetext[15]{This is in
contrast to the considered $\bar R^4-$model whose solutions $\bar R
(\phi)$ are unambiguously defined by the sign of the nonlinearity
parameter $\gamma$. The origin of this difference between the $\bar
R^{-1}-$ and the $\bar R^4-$model is in the number of real-valued
solutions $\bar R_i (\phi)$ of Eq. \rf{1.7} for given nonlinearity
parameters $\mu$ or $\gamma$ and $\phi$. In the case of the $\bar
R^{-1}-$model the corresponding quadratic equation has two
real-valued solutions for given $(\mu,\phi)$ whereas the cubic
equation \rf{4.2} has only one for fixed $(\gamma,\phi)$.} over
given values of the parameter $\mu$ and the nonlinearity field
$\phi$. At early evolution stages of the Universe, transitions
between these two branches cannot be ruled out a priori and should
be taken into account for a comprehensive description of the
dynamics of the Universe.

Finally, we note that the external spacetime in the considered pure
geometrical models is necessarily $AdS$ and the corresponding
negative effective cosmological constant, $\Lambda_{eff}<0$, forbids
a late-time acceleration. The situation can be cured by including
additional matter fields. Examples are flux field stabilization
scenarios \cite{GMZ2} which provide certain moduli space sectors
with positive effective cosmological constant and external
spacetimes of $dS$ type.

\bigskip
We thank Sugumi Kanno and Jiro Soda for useful discussions.
U.G. acknowledges support from DFG grant KON/1806/2004/GU/522. A.Z.
thanks the Physics Department of Universidade Federal de
Para$\acute{\mbox{\i}}$ba (Jo$\tilde{\mbox{a}}$o Pessoa, Brazil) for
their kind hospitality and CNPq for financial support. V.B.B. and
C.R. acknowledge partial financial support from CNPq and
CNPq/FAPESQ-Pronex.


\appendix
\section{The quartic equation and its associated quadratic equation sets\label{app1}}
In this appendix we briefly re-derive the quadratic equation sets
associated to the quartic equation \rf{4.4}. Following the Ferrari
formalism as it is briefly described, e.g., in \cite{Abramowitz} we
lay explicit emphasis on the sign rules\footnotetext[16]{These sign
rules are not displayed explicitly in \cite{Abramowitz}.} which are
crucial for a correct derivation of the final solution set of the
quartic equation.

The crux of the Ferrari formalism applied to a quartic equation of
type \rf{4.4},
\be{q1}
x^4+a_1 x +a_0=0,
\ee
consists in factoring it by transforming it into a difference of
two quadratic terms
\be{q2}
A^2-B^2=(A+B)(A-B)=0
\ee
so that solutions can be obtained from
\be{q2a}
A\pm B=0.
\ee
Adding and subtracting a term $x^2u+(u/2)^2$ in \rf{q1}, with $u$
an auxiliary function, one rewrites \rf{q1} as
\be{q3}
\left(x^2+\frac u2\right)^2-u\left(x^2-\frac{a_1}{u}x+\frac{\frac
14 u^2-a_0}{u}\right)=0
\ee
and requires the second term to be quadratic
\be{q4}
\left(x^2+\frac u2\right)^2-u\left(x+\epsilon \sqrt{\frac{\frac 14
u^2-a_0}{u}}\right)^2=0.
\ee
Here, $\epsilon$ is a sign factor $\epsilon =\pm 1$ and we assume
for definiteness
\be{q5}
\sqrt{\frac{\frac 14 u^2-a_0}{u}}>0 \quad \mbox{for} \quad
\frac{\frac 14 u^2-a_0}{u}>0.
\ee
The compatibility of Eqs. \rf{q3} and \rf{q4} is ensured by the
condition
\be{q6}
-\frac{a_1}{u}=2\epsilon\sqrt{\frac{\frac 14 u^2-a_0}{u}}
\ee
which on its turn is equivalent to the cubic equation
\be{q7}
u^3-4a_0u-a_1^2=0.
\ee
We specify as in \rf{4.4}, \rf{4.9}
\be{q9}
a_0:=-\frac{2D\Lambda_D}{\gamma(D-8)},\quad a_1:=\frac{D-2}{\gamma
(D-8)},\quad -4a_0=3q,\quad a_1^2=2r
\ee
and assume that $u$ is a real-valued solution of Eq. \rf{q7}. The
analysis of section \ref{dn8} shows that for stable configurations
of the $R^4-$model it holds additionally $u>0$. Using this
condition as simplifying input information, we can rewrite the
quartic equation \rf{q1}, \rf{q4} as
\be{q10}
\left(x^2+\frac{u}{2}\right)^2-\left(\sqrt{u}x+\frac{\epsilon}{2}
\sqrt{u^2+3q}\right)^2=0.
\ee
It factorizes according to \rf{q2}, \rf{q2a} into a set of
quadratic equations
\be{q11}
x^2\pm \sqrt{u}x+\frac 12\left(u\pm \epsilon
\sqrt{u^2+3q}\right)=0.
\ee
The sign factor $\epsilon$ follows from \rf{q5}, \rf{q6}, \rf{q9}
and $u>0$ as
\be{q12}
\epsilon=-\sign(a_1)=-\sign\left(\frac{D-2}{\gamma (D-8)}\right).
\ee

\section{Sign analysis of the discriminant $Q$\label{discrim}}
The sign of $Q$ can be obtained by mapping the minimum-ensuring
inequality \rf{4.4-1} into an equivalent inequality for $z$. For
this purpose we consider the critical surface $\Xi_{c1}\subset
{\cal M}$ in the parameter space, where for $\gamma\neq 0$ the
inequality \rf{4.4-1} is replaced by an equality,
\be{q8-1}
\Xi_{c1}=\left\{(\Lambda_D,\gamma,\bar R)\in {\cal M}| \quad
\xi_{c1}[\Lambda_D,\gamma,\bar R]:=(D-2)+4(D-8)\gamma
\bar R^{3}=0\right\}.
\ee
The intersection of this surface $\Xi_{c1}$ with the algebraic
variety ${\cal V}:\ h[\Lambda_D,\gamma,\bar R]=0$ of the extremum
condition will define a critical value $z_{c1}$. This value can be
found explicitly by resolving \rf{q8-1} for $\bar R$, what gives
\be{q8-2}
\left.\bar
R_{c1}:=R\right|_{\Xi_{c1}}=-\left(\frac{D-2}{4(D-8)\gamma}\right)^{1/3},
\ee
and plugging  $\bar R_{c1}$ into the quartic equation \rf{4.4}. As
result one obtains
\be{q8-3}
z_{c1}(\Lambda_D,\gamma )=4\gamma (8\Lambda_D/3)^3 w(D)=-1\, .
\ee
Now, small perturbations off the critical surface $\Xi_{c1}$, but
along the variety ${\cal V}$, can be used to map the inequality
\rf{4.4-1} into its counterpart for $z$. Setting
\be{q8-4}
\Lambda_D=\Lambda_{D,c1}+\delta\Lambda_D, \qquad \bar R= \bar
R_{c1}+\delta \bar R, \qquad z=z_{c1}+\delta z
\ee
and keeping $\gamma$ fixed, we get from the inequality \rf{4.4-1}
\be{q8-5}
72(D-8)\gamma^2 \bar R_{c1}^2\delta \bar R>0
\ee
whereas the quartic equation \rf{4.4} and the definition \rf{q8}
of $z$ yield
\be{q8-6}
\delta\Lambda_D=3\gamma\frac{D-8}{D}\bar R_{c1}^2 \left( \delta
\bar R\right)^2, \qquad \delta z=12 \gamma w(D)(8/3)^3
\Lambda_{D,c1}^2\delta\Lambda_D
\ee
and hence
\be{q8-7}
\delta z= \frac{2^{11}}{3}\frac{D^2(D-8)^2}{(D-2)^4}\left(\gamma
\bar R_{c1} \Lambda_{D,c1}\right)^2\left( \delta \bar R\right)^2\ge 0\, .
\ee
We observe that, although inequality \rf{q8-5} implies
\be{q8-8}
D>8: \  \delta \bar R>0,\qquad D<8: \ \delta \bar R<0\, ,
\ee
independently of the signs of $\gamma$ and $\bar R_{c1}$, the
variety ${\cal V}$ is for $\gamma\neq 0$, $D\neq 8$ and (because
of $\delta z>0$) located over the region
\be{q8-9}
z(\Lambda_D,\gamma )> z_{c1}(\Lambda_D,\gamma )=-1
\ee
of the $(\Lambda_D,\gamma )-$plane. This means that by any
perturbation (motion) on the variety ${\cal V}$ we cannot pass
across the critical value $z_{c1}(\Lambda_D,\gamma )=-1$. Hence,
$z_{c1}(\Lambda_D,\gamma )=-1$ must be a boundary segment of the
projection $\pi {\cal V}$ of ${\cal V}$ onto the
$(\Lambda_D,\gamma)-$plane:  $z_{c1} \subset
\partial \left(\pi {\cal V}\right)$. The latter fact is confirmed by
the observation that the critical surface $\Xi_{c1}\subset {\cal
M}$ coincides with the singular surface
$$\partial_{\bar R} h[\Lambda_D,\gamma,\bar R]=(D-2)+4(D-8)\gamma
\bar R^{3}=0
$$
of the projection $\pi$ of ${\cal V}$ onto the
$(\Lambda_D,\gamma)-$plane\footnotetext[17]{Singularities of smooth
projections are extensively discussed, e.g., in Refs.
\cite{arnold}.}.

\section{Mapping $f'>0$ into parameter space\label{phys-branch}}
The inequality $f'=1 + 4\gamma \bar R^3>0$ can be analyzed with
the same technique as the minimum-ensuring inequality \rf{4.4-1}
(see relations \rf{q8-1} - \rf{q8-9}): we obtain the intersection
of the critical surface
\be{4.11-7}
\Xi_{c2}=\left\{(\Lambda_D,\gamma,\bar R)\in {\cal M}| \quad
\xi_{c2}[\Lambda_D,\gamma,\bar R]=1 + 4\gamma \bar R^3=0\right\}
\ee
with the algebraic variety ${\cal V}$, i.e. $\Xi_{c2}\cap {\cal
V}$, and study the behavior of small parameter perturbations off
$\Xi_{c2}$ and along the variety ${\cal V}$.

Explicitly this means that we resolve \rf{4.11-7} for $\bar R$ to
obtain
\be{4.11-8}
\bar R_{c2}=-(4\gamma)^{-1/3}
\ee
and plug this $\bar R_{c2}$ into the quartic equation \rf{4.4}.
{}From the intermediate result
\be{4.11-9}
4\gamma(8\Lambda_D/3)^3=-1
\ee
we find by multiplication with $w(D)$ that the intersection
$\Xi_{c2}\cap {\cal V}$ corresponds to the critical value
\be{4.11-10}
z_{c2}(\Lambda_D,\gamma)=-w(D)\, .
\ee
Substituting, furthermore, the perturbation ansatz
\ba{4.11-11}
\Lambda_D&=&\Lambda_{D,c2}+\delta\Lambda_D, \qquad \bar R= \bar
R_{c2}+\delta \bar R, \nn\\ z&=&z_{c2}+\delta z,\qquad f'=\delta f',
\ea
(it holds $f'_{c2}=0$) into the defining relation \rf{q8} for $z$,
the quartic equation \rf{4.4}, and the equality  $f'=1+4\gamma
\bar R^3$,  we get for fixed $\gamma$
\ba{4.11-12}
\delta z&=&12\gamma w(D) (8/3)^3\Lambda_{D,c2}^2\delta\Lambda_D\,
,\qquad \delta\Lambda_D=\frac 3D \delta \bar R\, , \nn\\ \delta
f'&=&12\gamma \bar R_{c2}^2\delta \bar R\, ,
\ea
respectively, and by combination of these results also
\be{4.11-13}
\delta\Lambda_D=\frac{1}{4\bar R^2 D\gamma}\delta f',\qquad \delta
z=\frac 3D\left(\frac 83\right)^3\left(\frac{\Lambda_{D,c2}}{\bar
R_{c2}}\right)^2w(D)\delta f'.
\ee
{}From the definition \rf{q8} of $w(D)$ and its implication
\be{4.11-14}
w(D<8)<0,\qquad 0<w(D>8)<1
\ee
we find\footnotetext[18]{\label{f-}For completeness we note that a
formal crossing of the critical surface $\Xi_{c2}$ into the sector
$f'<0$ would lead to the restrictions
$$
D<8:\ z>|w(D)|,\qquad D>8:\ z<-w(D).
$$
The sector $f'<0$ itself would correspond to a negative effective
gravitational constant in BD frame (an antigravity sector)
\cite{antigrav1,antigrav2}. Antigravity effects are described for
various types of higher dimensional setups. In certain SUGRA
configurations, spatially bounded antigravity regions are known as
repulsons (white holes) \cite{repulson}. Several brane world
models show radion induced scalar antigravity at ultra-large
distances \cite{antigrav3}. In our model, the antigravity sector
would fill a complete external spacetime which necessarily would
be disconnected from our own observable Universe with $f'>0$ by a
conformal singularity at $f'=0$ (see subsection \ref{stab} below
and also \cite{antigrav2}).  There remains the interesting open
question of whether the formal correspondence between the
strong-curvature sector of the $\bar R-$nonlinear model and the
emerging antigravity sector of the $R-$linear model in BD-frame
simply signals an inconsistency of the theory and whether it can
find a physically meaningful interpretation in an enlarged
(extended) setup. The repulson in SUGRA models was reinterpreted
as unphysical region and found its resolution by the enhan\c{c}on
mechanisms (excision of the antigravity region and placing a heavy
shell of wrapped D-branes on its boundary so that the former
antigravity region in the interior is shielded by the D-brane
source and replaced by a segment of flat space) \cite{sing}. In
analogy to the enhan\c{c}on mechanism in the repulson case, one
can expect a resolution of the conformal singularity at $f'=0$ by
some kind of quantum gravity mechanism.} for $\delta f'>0$, $f'>0$
\ba{4.11-15an}
D<8:& \qquad & z_{c2}(\Lambda_D,\gamma)=|w(D)|>0, \qquad
z<-w(D)=|w(D)|\label{4.11-15aa}\, ,\\
D>8:& \qquad & z_{c2}(\Lambda_D,\gamma)=-w(D)<0, \qquad
-w(D)<z\label{4.11-15ba}\, .
\ea

\section{Parameter limits\label{limits}}
In subsection \ref{stab} it has been found that for the
$(f'>0)-$sector the boundary segments
$z(\Lambda_D,\gamma)=-w(D)\subset \partial
\Theta_{(\Lambda_D,\gamma)}$ of the projection
$\Theta_{(\Lambda_D,\gamma)}:=\pi \Upsilon$ of the stability region
$\Upsilon\subset {\cal V}\subset {\cal M}$ onto the
$(\Lambda_D,\gamma)-$plane correspond to the limit $\phi\to-\infty$.
Here, we clarify the behavior of the system in the vicinity of the
other boundary segments $\partial \Theta_{(\Lambda_D,\gamma)}
\supset \{\Lambda_D=0\cup \gamma=0\}$ for $D>8$ (see Fig.
\ref{fig2b}) and $\partial \Theta_{(\Lambda_D,\gamma)} \supset
\{\Lambda_D=0\cup \gamma=0\cup z=-1\}$ for $D<8$ (see Fig.
\ref{fig2a}).\\

\noindent {\boldmath $\Lambda_D\to 0, \gamma\neq 0:$} In this limit
we obtain from Eqs. \rf{4.9}, \rf{q8}, \rf{4.11}
\be{l1}
q\to 0, \quad z\to 0,\quad Q\to r^2\neq 0, \quad v_1 \to 2^{1/3}
\ee
and hence from Eqs. \rf{4.11-6}, \rf{4.11-6a}, \rf{e1}
\ba{l2}
\gamma>0:&\qquad &\bar R(\Lambda_D\to -0)\to -0\quad \Longleftarrow
\quad \left\{\begin{array}{lcl}
  \bar R_{-,+}&\to &-0 \\
  \bar R&\to &-0 \\
  \bar R_{+,-}&\to &-0 \\
\end{array}\qquad\begin{array}{c}
  \mbox{for}\quad D>8,  \\
  \mbox{for}\quad D=8, \\
  \mbox{for}\quad D<8, \\
\end{array}\right.\label{l2a}\\
 \gamma<0:&\qquad &\bar R_{-,-}(\Lambda_D\to 0)\to
-(2r)^{1/6}\qquad \mbox{for}\quad D<8.\label{l2b}
\ea
Obviously, the system behaves differently in the upper and lower
$(\Lambda_D,\gamma)-$plane. In the case of $\gamma<0$, the system
behaves regularly for $\Lambda_D \to 0$ and the half-line
$(\Lambda_D=0,\gamma<0)$ is not distinguished from its vicinity. In
contrast to this, the limit $(\Lambda_D\to -0,\gamma>0)$ corresponds
to a flat-space limit $\bar R\to -0$ which via \rf{4.2} is
associated with a freezing of the nonlinearity field $\phi$:\ \
$f'\to 1$ at $\phi_0\to 0$. From  Eq. \rf{1.11b} follows
\be{12-2}
\left.\partial_\phi^2 U\right|_{\phi_0}\approx
\frac{D-2}{24(D-1)}\frac{1}{\gamma \bar R^2}\to +\infty
\ee
so that for the mass of the nonlinearity field holds $m_\phi^2\to
+\infty$. We note that the nonlinearity field in an $R^2-$model has
a finite mass $m_\phi$ in the limit $\Lambda_D \to -0$  (see, e.g.,
\cite{GMZ1,GMZ2}). The different behavior of the models is caused by
the different powers of the term $(e^{A\phi} - 1)$ in $U(\phi )$:
for an $R^2-$model this power equals 2, whereas for an $R^4-$model
it equals 4/3. Hence, in the latter case the second derivative
$\left. d^2 U(\phi )/d\phi^2\right|_{\phi_{(0)}}$ diverges in the
limit $\phi_{(0)} \to 0$.

We arrived at the interesting fact that in the considered toy model
the extremum condition in form of the quartic equation \rf{4.4}
relates the scalar curvature $\bar R$ at the minimum and the bare
cosmological constant $\Lambda_D$ in the case of $\gamma>0$ so
strongly that for stabilized internal spaces the limit $\Lambda_D\to
0$ corresponds to the flat-space limit $\bar R\to -0$. As it should
be, the flat-space limit of the total scalar curvature $\bar R\to
-0$ implies via \rf{1.12-1}, i.e. $U(\phi_0)\to -0$, and \rf{1.36},
$\hat R_i=2d_i U(\phi_0)/(D-2)$, also a decompactification of the
internal space components $\hat R_i=e^{-2\beta^i_0}R_i\to -0$,
$\beta^i_0\to
+\infty$ (the $R_i$ are hold fixed).\\

\noindent {\boldmath $\gamma\to 0, \Lambda_D\neq 0:$} The
definitions \rf{4.9} and \rf{q8} show that for fixed $\Lambda_D\neq
0$ the limit $\gamma\to 0$ implies
\be{l3}
r,|q|,Q\to +\infty, \qquad z\to 0.
\ee
With the help of an expansion in terms of small $z\approx 0$ the
rescaled curvatures \rf{4.11-6a} are easily obtained from
\rf{4.11-0} as
\ba{lim10}
T_{+,-}(z\to 0)&\approx &-3\times 2^{-5/2}z^{1/3}\nn\\
T_{-,+}(z\to 0)&\approx &3\times 2^{-5/2}z^{1/3}\nn\\
T_{-,-}(z\to 0)&\approx &-2^{1/6}- 2^{-5/2}z^{1/3}
\ea
so that the curvatures $\bar R_{\epsilon,\pm}$ themselves can be
estimated via $\bar R_{\epsilon,\pm}=r^{1/6}T_{\epsilon,\pm}$
as\footnotetext[19]{The general limiting behavior $\bar R(\gamma\to
0)$ without identification of the concrete solution branch $\bar
R_{\epsilon,\pm}$ can be easily obtained from the quartic equation
\rf{4.4}. Assuming $\bar R (\gamma\to 0)<\infty$ and taking the
limit $\gamma\to 0$ in Eq. \rf{4.4} gives $\bar
R=2D\Lambda_D/(D-2)$, whereas division of \rf{4.4} by $\bar R$ for a
behavior $|\bar R(\gamma\to 0)|\to \infty$ yields $\bar
R=-\left(\frac{D-2}{\gamma (D-8)}\right)^{1/3}$.}
\ba{lim11}
\left.\begin{array}{c}
  \bar R_{-,+}(\gamma\to +0;D>8) \\
  \bar R_{+,-}(\gamma\to +0;D<8) \\
\end{array}\right\}&\approx &
\frac{2D\Lambda_D}{D-2},\label{lim11a}\\
\bar R_{-,-}(\gamma\to -0;D<8)&\approx &-\left|\frac{D-2}{\gamma
(D-8)}\right|^{1/3}\to -\infty\, .\label{lim11b}
\ea
In the exceptional $(D=8)$ case the scalar curvature in the minimum
does not depend on $\gamma$ and is given by Eq. \rf{e1}
\be{lim11-2}
\bar R=\frac 83 \Lambda_D=\frac{2D\Lambda_D}{D-2}\, .
\ee

Again, the system behaves qualitatively different in the upper and
the lower $(\Lambda_D,\gamma)-$plane. Because of the finite
asymptotics \rf{lim11a} and Eq. \rf{lim11-2}, in the upper
half-plane it holds (for $D<8$ and $D\ge 8$)
\be{lim12}
f'(\gamma\to+0)\to 1,\quad \phi_0\to 0
\ee
\ba{lim13}
\left.\partial_\phi^2U(\gamma\to +0)\right|_{\phi_0}&\approx&
\frac{(D-2)^3}{96(D-1)D^2}\frac{1}{\gamma\Lambda_D^2}\to
+\infty\label{lim13a}\\
U(\phi_0;\gamma\to +0)&\to & \Lambda_D\label{lim13b}
\ea
and the nonlinearity field $\phi$ undergoes a freezing stabilization
at $\phi_0=0$ with diverging mass $m_\phi \to +\infty$ but finite
scalar curvature \rf{lim11a} and finite minimum position
$U(\phi_0).$ Hence, under the freezing stabilization of the
nonlinearity field for $\gamma\to+0$ the system turns smoothly into
a system with linear scalar curvature term $\bar R$, i.e. into a
system with Einstein-Hilbert action in $\bar R$. This is a generic
feature of models with nonlinear scalar curvature terms and was
earlier described for $R^2-$models in \cite{GMZ1,GMZ2}. Figure
\ref{fig3} gives a rough illustration of the corresponding
deformation of the potential $U(\phi)$ under variation of $\gamma$
and for a fixed value of $\Lambda_D$.

The behavior of the system is completely different in the
lower-half-plane limit $\gamma\to -0$. Here we have to distinguish
the dimensions $D=8$ and $D<8$. For $D=8$ Eqs. \rf{lim13a},
\rf{lim13b} extend to the lower half-plane $\gamma\to -0$, i.e. the
system is completely unstable in this limit
$\partial_\phi^2U|_{\phi_0}(\gamma\to -0) \to -\infty$. Obviously,
$\partial_\phi^2U|_{\phi_0}\sim 1/\gamma$ in \rf{lim13a} encounters
a pole singularity with respect to $\gamma$. This is different for
$\bar R_{-,-},\ D<8$. Here, one finds from \rf{4.2}, \rf{lim11b}
\be{lim14}
e^{A\phi_0}=f'(\gamma\to -0)\to \frac{3D}{|D-8|}
\ee
and from \rf{1.11b}, \rf{1.12-1}
\ba{lim15}
\left.\partial_\phi^2U(\gamma\to -0)\right|_{\phi_0}&\approx&
-\frac{1}{\gamma^{1/3}}\frac{(D-2)^{1/3}(D-8)^{2/3}}{8(D-1)}\left|\frac{3D}{D-8}\right|^{-\frac{2}{D-2}}\to
+\infty\label{lim15a}\\  U(\phi_0;\gamma\to -0)&\approx &
\frac{1}{\gamma^{1/3}}\frac{(D-2)^{4/3}}{2D|D-8|)}\left|\frac{3D}{D-8}\right|^{-\frac{2}{D-2}}\to
-\infty\label{lim15b}.
\ea
From these equations and their rough illustration in Fig. \ref{fig4}
we read off that in the limit $\gamma\to -0$ the minimum of the
potential $U(\phi)$ lowers infinitely, $U(\phi_0;\gamma\to -0)\to
-\infty$, and becomes fixed at the finite value $\phi_0=\frac 1A \ln
\left|\frac{3D}{D-8}\right|$. The corresponding scalar curvature
diverges as $\bar R_{-,-}(\gamma\to -0)\to -\infty$ and is separated
from the conformal singularity $f'(\phi\to -\infty)\to 0$ by a
barrier whose top is defined by the associated maximum branch $\bar
R_{-,+}$ and tends to the value
\be{lim15-2}
\bar R_{-,+}(\gamma \to -0)\to \frac{2D\Lambda_D}{D-2}, \qquad
\left.U\right|_{max}(\gamma\to -0)\to \Lambda_D
\ee
(see also Fig. \ref{fig4}). This means that the metastable sector in
the lower $(\Lambda_D,\gamma)-$plane is separated from absolutely
stable systems in the upper half-plane and their limiting linear
$(\gamma\to +0)-$models by an infinite gap. Considering the behavior
of the system over the $(\Lambda_D,\gamma)-$plane we have to
conclude that it possess an essential singularity in the limit
$\gamma\to -0$, i.e. the scalar curvature $\bar R$ encounters an
infinite jump between $\gamma\to +0$ $(\bar R\to 2D\Lambda_D/(D-2))$
and $\gamma\to -0$ $(\bar R\to -\infty)$ whereas $\partial_\phi^2
U|_{\phi_0}$ has for $\gamma>0$ a pole-like singularity $\sim
\gamma^{-1}$ and for $\gamma<0$ it behaves like a branching  point
singularity $\sim \gamma ^{-1/3}$. This is a strong indication that
the description of a physical system in terms of the considered
oversimplified toy model breaks down in the limit $\gamma\to -0$.
The search for possibly existing physically realistic metastable
systems with $\gamma<0$ is out of the scope of the present work and
we leave a corresponding
investigation to future research.\\

\noindent {\boldmath $z\to -1:$} Relations \rf{4.11-18a} -
\rf{4.11-18d} show that this limit can only be reached by metastable
configurations $\gamma<0,\ D<8, \ \bar R_{-,-}$. According to
\rf{q8} it corresponds to a vanishing discriminant $Q=0$ of the
cubic equation \rf{4.8} so that this equation has two coinciding
solutions $u_{1,2}=2r^{1/3}$  (it holds $v_{1,2}(z=-1)=2$). From
Eqs. \rf{4.11-6}, \rf{4.11-6a} we observe that this leads to a
coalescence of the minimum branch $\bar R_{-,-}$ and the associated
maximum branch $\bar R_{-,+}$ in an inflection point at $z=-1$
\be{lim16}
\bar R_{-,\pm}(z=-1)=-\frac 12
\sqrt{u_1(z=-1)}=-\frac{r^{1/6}}{\sqrt
2}=-2^{-2/3}\left(\frac{D-2}{\gamma (D-8)}\right)^{1/3}.
\ee
The situation is also obvious from Fig. \ref{fig1a}. Configurations
with $z\le -1$ have no extremum at all and are necessarily unstable.

\end{document}